\documentclass[a4paper,12pt]{article}

\usepackage{amsfonts}
\usepackage{amsmath}
\usepackage{graphicx,epsfig}
\usepackage{titlesec}
\numberwithin{equation}{section}

\begin{document}

\makeatletter

%%%%% Unnecessary for REVTEX %%%%%%
\titleformat{\section}
  {\large\bfseries}{\thesection.}{0.5em}{}
\titleformat{\subsection}
  {\normalfont\bfseries}{\thesubsection.}{0.5em}{}

%%%%% Caption style %%%%%
\renewcommand{\fnum@figure}{\textbf{Fig.~\thefigure}}
\renewcommand{\fnum@table}{\textbf{Table~\thetable}}

\newcommand{\beq}{\begin{equation}}
\newcommand{\eeq}{\end{equation}}
\newcommand{\la}{\langle}
\newcommand{\ra}{\rangle}
\newcommand{\bu}{\bullet}
\newcommand{\wt}{\widetilde}
\newcommand{\wh}{\widehat}
\newcommand{\ol}{\overline}
\newcommand{\nn}{\nonumber}
\newcommand{\ob}{\overbrace}
\newcommand{\ub}{\underbrace}

\makeatother

\allowdisplaybreaks[1]

\begin{titlepage}

%%%%% Footnote symbol in title page %%%%%
\renewcommand{\thefootnote}{\fnsymbol{footnote}}

%%%%% Preprint number %%%%%
\hfill\parbox{4cm}{hep-th/0607075} 

%%%%% Title %%%%%
\vspace{35mm}
\baselineskip 9mm
\begin{center}
{\LARGE \bf Recursive Relations \\ for a Quiver Gauge Theory}
\end{center}

%%%%% Authors and Addresses %%%%%
\baselineskip 6mm
\vspace{10mm}
\begin{center}
Jaemo Park\footnote{\tt jaemo@postech.ac.kr}
and Woojoo Sim\footnote{\tt space@postech.ac.kr} 
\\[5mm] 
{\sl Department of Physics, POSTECH, Pohang 790-784, Korea \\} 
\end{center}

%%%%% Abstract %%%%%
%\vfill
\vspace{20mm}
\begin{center}
{\bf Abstract}
\end{center}
\indent 
We study the recursive relations for a quiver gauge theory with the gauge group $SU(N_1)\times SU(N_2)$ with bifundamental fermions transforming as $(N_1,\bar{N_2})$. 
We work out the recursive relation for the amplitudes involving a pair of quark and antiquark and gluons of each gauge group. 
We realize directly in the recursive relations the invariance under the order preserving permutations of the gluons of the first and the second gauge group.
We check the proposed relations for MHV, 6-point and 7-point amplitudes and find the agreements with the known results and the known relations with the single gauge group amplitudes.
The proposed recursive relation is much more efficient in calculating the amplitudes than using the known relations with the amplitudes of the single gauge group.

\end{titlepage}

%%%%% Footnote symbol in body %%%%%
\renewcommand{\thefootnote}{\arabic{footnote}}
\setcounter{footnote}{0}

\section{Introduction}
Recently, we have witnessed the rapid progress in the calculation of the multiparton amplitudes.
The inflection point is the seminal work by Witten \cite{witten0312}, a twistor theoretical reformulation of perturbative super Yang-Mills theory in terms of topological string theory.
Witten established a correspondence between multiparticle amplitude in $\mathcal{N}=4$ super Yang-Mills on $\mathbb{R}^{3,1}$ 
and open string amplitude in the topological $B$ model in the Calabi-Yau super manifold $\mathbb{CP}^{3|4}$.
Maximally helicity-violating (MHV) amplitudes at tree level were successfully reproduced.
Motivated by this, subsequent progresses were made to give a prescription for computing MHV and non-MHV amplitudes at the tree level and beyond.
\\ \indent
Two important developments are worth mentioning, related to the current work. 
One is the work done by Cachazo, Svrcek and Witten (CSW) \cite{csw0403}, where the simple rules were suggested for computing non-MHV amplitudes using the MHV amplitudes as vertices and a simple off-shell description for the propagator.
This method is extended to one-loop computations in \cite{bst0407}\cite{qr0410}\cite{bbst0410}\cite{bbst0412}\cite{csw0409}\cite{bddk9403}\cite{bddk9409}\cite{bdk0412}.
\\ \indent
The other is the development of the BCFW recursive relation \cite{bcf0412}\cite{bcfw0501}, where one can find the recursive relation between an $n$-point gluon amplitude and amplitudes of smaller number of external particles.
Thus if we know just $3$-point gluon amplitudes, general $n$-point amplitudes can be worked out recursively.
This recursion relation at the tree level is extended to the case involving fermions in \cite{lw0501}\cite{lw0502}. 
The extension of the recursion relation to the one-loop computations is subsequently developed  \cite{bdk0501}\cite{bdk0505}\cite{bdk0507}\cite{bbdi0507}.
\\ \indent
In this paper, we work out the tree level recursion relation and CSW rules for a quiver gauge theory $SU(N_1)\times SU(N_2)$ with bifundamental fermions transforming as $(N_1,\bar{N_2})$.
Given that such quiver gauge theories are dual to the topological string on suitable super orbifolds \cite{pr0411}, this is a natural generalization of the recursive relation of a single gauge group $SU(N)$, which can be regarded as a subamplitude of $\mathcal{N}=4$ super Yang-Mills theory.
In addition, this work is relevant for the computation of multiparticle amplitudes of the standard model in a high energy regime well above the Fermi scale \cite{dy91}\cite{my93}\cite{mahlon93}.
In \cite{witten0312} the MHV amplitudes were worked out using the topological string theory.
Here we work out more general amplitudes using the recursive relations.
Furthermore, CSW relation is suggested and is checked for the simplest cases.
\\ \indent
The resulting recursive relation is easy to state.
We are mostly interested in amplitudes involving one bifundamental pairs $q\bar{q}$ and gluons of two gauge groups.
The relevant subamplitudes can be obtained by summing over subamplitudes of a gauge theory of a single factor with order-preserving reshuffling of the gluons of two different gauge groups.
The recursive relation of the product gauge group can be obtained by taking care of such reshuffling at the level of recursive relation directly.
\\ \indent
If we have amplitudes invariant under such order preserving reshuffling up to $n$-point amplitudes, the $(n+1)$-point amplitudes can be obtained using the similar recursive procedure to a single gauge group case together with the addition of the contributions generated by such reshuffling.
The CSW relation can be obtained using the MHV amplitudes for the product gauge group as vertices and the usual off-shell prescription for the propagator.
\\ \indent
The contents of this paper are as follows.
In section \ref{bcfw}, we review the basic spinor formalism and the BCFW recursive relations. In section \ref{rr} we provide relevant background and work out some examples of the recursive relations for the product gauge group. Section \ref{mhv}, \ref{6p} and appendix \ref{7p} provide detailed calculation results for MHV, 6-point, and 7-point amplitudes respectively.
In section \ref{csw}, finally, we work out the CSW relations. Several technical details are relegated to appendices.
\\ \indent
Some of our conventions are as follows.
\begin{enumerate}\setlength\itemsep{-\parsep}
\item Particle 1 and 2 denote $+$ helicity antiquark and $-$ helicity quark respectively. 
                        We take all external momenta to be outgoing.
\item Particle 3, 4, $\cdots$ denote gluons.
      (For the convenience, gluons of different gauge groups are \emph{not} distinguished by notations such as bars.)
\item $A_{ab\cdots}^{\cdots d^+\cdots f^-\cdots}$ denotes a subamplitude of the product gauge group theory. 
Here the superscripts denote all gluons and their helicities and the subscripts denote the gluons which belong to the first gauge group. 
For example, $A_{3}^{4^+5^+3^-6^-}$ denotes a 6-point subamplitude where gluon $3^-$ belongs to the first gauge group and $4^+$, $5^+$, $6^-$ to the second.
We omitted fermions since in most cases they are $1^+$ and $2^-$. If not the case, we specify them with extra superscripts in front of the gluons as $A_{ab\cdots}^{(\ol{q}^+q^-)\cdots d^+\cdots f^-\cdots}$. 
e.g. $A_{5}^{(P^+2^-)5^+6^-}$ denotes a 4-point subamplitude where $P^+$ and $2^-$ are fermions and $5^+$ and $6^-$ are gluons of the first and the second gauge group respectively. 
Finally, if the gluons are in numerical order, we omit the gluon indices and indicate only their helicities as  
$A_{ab\cdots}^{\cdots ^+\cdots ^-\cdots}$. e.g. $A_{34}^{--+++}$ denotes a 7-point subamplitude where gluon $3^-$ and $4^-$ belong to the first gauge group and $5^+$, $6^+$, $7^+$ to the second.
\item $A_S^{\cdots}$ denotes a subamplitude of the single gauge group theory and $A_G^{\cdots}$ a purely gluonic subamplitude.
The superscripts are the same as the cases of the product gauge group theory.
\item When there is no confusion, the term \emph{amplitude} is used in replacement for the term \emph{subamplitude}.
\end{enumerate}

\section{BCFW Recursive Relations} \label{bcfw}

For the momentum $p_{\mu}$ of a massless particle in four dimension, one can use the spinor representation as,  $p^{a\dot{a}}=p^{\mu}\sigma_{\mu}^{a\dot{a}}=\lambda_p^a \wt{\lambda}_p^{\dot{a}}$ where $\lambda_p^a$ and $\wt{\lambda}_p^{\dot{a}}$ are positive and negative chirality spinors respectively \cite{bg87}.
(On Minkowskian, $\wt{\lambda}_p^{\dot{a}}=\ol{\lambda_p^a}$.)
Moreover, the on-shell states of the particles in a scattering process, spinors and helicity vectors, can be expressed in terms of these momentum spinors. 
In result, a subamplitude of the scattering process is expressed in terms of only spinorial inner products, which are defined as $\la ij\ra=\epsilon_{ab}\lambda_{p_i}^a\lambda_{p_j}^b$ and $[ij]=\epsilon_{\dot{a}\dot{b}}\lambda_{p_i}^{\dot{a}}\lambda_{p_j}^{\dot{b}}$.
\\ \indent
Before mentioning the BCFW recursion relation, let us list some notations used for the spinor expressions of the amplitudes \cite{witten0312}\cite{bcf0412}.
\beq
\begin{split}
t_{ijk\cdots} &\equiv (p_i+p_j+p_k+\cdots)^2 \\
\la i|\sum_r p_r|j] &\equiv \sum_r \la ir\ra[rj] \\
\la i|(\sum_r p_r)(\sum_s p_s)|j\ra &\equiv \sum_r \sum_s \la ir\ra[rs]\la sj\ra \\
\quad [i|(\sum_r p_r)(\sum_s p_s)|j] &\equiv \sum_r \sum_s [ir]\la rs\ra[sj] \\ 
\end{split}
\eeq
\indent
Now, the BCFW recursion relation states that a subamplitude is expressed as a sum over the products of two subamplitudes of smaller number of particles and a scalar propagator as,
\beq \label{eq:rr}
\mathcal{A}_n(\cdots,\wh{m},\cdots,\wh{n},\cdots)
= \sum_{S}\sum_{h=+,-}\mathcal{A}(\cdots,\wh{m},\cdots,\wh{P}^h)
                \frac{1}{P^2}\mathcal{A}(-\wh{P}^{-h}\cdots,\wh{n},\cdots)
\eeq
where 
\begin{align} \label{eq:tool1}
%\begin{split}
P &= \sum_{part} p_i, \nn \\ 
\wh{P} &= P + \frac{P^2}{\la m|P|n]}\lambda_m\wt{\lambda}_{n}, \\
\wh{p}_m &= p_m + \frac{P^2}{\la m|P|n]}\lambda_m\wt{\lambda}_{n}, \nn \\
\wh{p}_n &= p_n - \frac{P^2}{\la m|P|n]}\lambda_m\wt{\lambda}_{n}  \nn
%\end{split}
\end{align}
\indent
In (\ref{eq:rr}), the sum $S$ is over all possible decompositions of external particles into two parts, keeping the overall orders of the particles determined according to the color factorization.\footnote{In purely gluonic cases and single gauge group cases there is only one overall order, but in product gauge group cases there exist several different overall orders generated by so called OPP(order preserving permutation), which is explained in section \ref{rr}.} On decomposition, $\mathcal{A}(\cdots,\wh{m},\cdots,\wh{P}^h)$, which we call a upper amplitude, is the amplitude of the particles of the first part and $\wh{P}$, whereas  $\mathcal{A}(-\wh{P}^{-h}\cdots,\wh{n},\cdots)$, a lower amplitude, is the amplitude of the particles of the second part and $-\wh{P}$.
We take, by convention, the sum in the first equation of (\ref{eq:tool1}) to be over all particles of the second part. 
In the recursive process, the reference particles $\wh{m}$ and $\wh{n}$ stay in the upper and lower amplitudes respectively.
\\ \indent
We will call each element in the recursive sum a \emph{configuration} and denote it as (particles in the upper amplitude $|$ $h,-h$ $|$ particles in the lower amplitude). 
For example, $(5^+6^-1^+\wh{2}^-|^{+-}|\wh{3}^+4^-)$ denotes $\mathcal{A}(5^+,6^-,1^+,\wh{2}^-,\wh{P}^+)
                \frac{1}{P^2}\mathcal{A}(-\wh{P}^{-},\wh{3}^+,4^-)$
where $P=3+4$.

\section{Recursive Relations for the Quiver Gauge Theory} \label{rr}

Consider a quiver type gauge theory with product gauge group $SU(N_1)\times SU(N_2)$.
Consider also quarks which transform in the bifundamental representation $(N_1, \bar{N_2})$ under these gauge groups and their complex conjugates. 
Then the full multiparton amplitudes $\mathcal{M}$ involving a quark-antiquark pair with $n_1$ gluons of $SU(N_1)$ and $\wt{n}_2$ gluons of $SU(N_2)$ can be written as \cite{mp91}\cite{pr0411},
\beq \label{eq:total}
\begin{split}
\mathcal{M}(q,1,\cdots,n_1 ; \wt{1},\cdots,\wt{n}_2,\bar{q})
&= \sum_{P(n_1),P(n_2)}(X^1\cdots X^{n_1})_{ij}(Y^1\cdots Y^{n_2})_{\bar{j}\bar{i}} \\
& \quad \times \mathcal{A}_{N_1 N_2}(q,1,\cdots,n_1,\bar{q} ; q,\wt{1},\cdots,\wt{n}_2,\bar{q})
\end{split}
\eeq
Here the sum is over all permutations of $n_1$ gluons of $SU(N_1)$ gauge group between the quark and the antiquark and similarly of $\wt{n}_2$ gluons of $SU(N_2)$ gauge group between the antiquark and the quark.
The $ij$, $\bar{j}\bar{i}$ indices refer to $SU(N_1)$ and $SU(N_2)$ color indices at the quark-antiquark pair.
$X^A$, $Y^B$ are generators of $SU(N_1)$ and $SU(N_2)$ gauge groups, respectively.
\\ \indent
Eq. (\ref{eq:total}) implies that the subamplitude $\mathcal{A}_{N_1 N_2}$ is invariant under the all possible permutations of the gluons maintaining the order of both the first and the second set of gluons. 
We call this permutation the order preserving permutation(OPP).
The invariance of the subamplitude under OPP is due to the fact that all Feynman diagrams with such permutations of the gluons have the same color factor $(X^1\cdots X^{n_1})_{ij}(Y^1\cdots Y^{n_2})_{\bar{j}\bar{i}}$ and contribute to the same $\mathcal{A}_{N_1 N_2}$.
\\ \indent
Though in the Feynman diagrams we should sum over all diagrams generated by OPP, the situation can be greatly improved if we encode the invariance under OPP directly for the BCFW type recursive relation.
In our formalism, the amplitudes which are related by OPP are equivalent.
\\ \indent
Suppose this type of equivalence holds up to $n$-point amplitudes. 
If we consider $(n+1)$-point amplitudes, this could be obtained from the $n$-point amplitudes and lower ones using the recursive relation similar to the single gauge group case. 
The result is not invariant under OPP in general.
We apply OPP again to the resulting configurations.
Many of the OPP elements do not generate new amplitudes.
We call an OPP \emph{reducible}, which does not exchange the particles in the upper amplitude with those in the lower amplitude in the expression appearing in the recursive relation.
Such a reducible OPP does not generate anything new, since the upper and the lower amplitudes are separately invariant under the OPP.
Only an irreducible OPP, which is not a reducible OPP, can generate new configurations.
For an $(n+1)$-point amplitude, we just sum over all such configurations generated by irreducible OPP.
\\ \indent
Also we know that in the Feynman diagram there are no vertices connecting gluons of different gauge groups in the product gauge group theory.
This is translated to the rule in the recursive relation for the product gauge group that whenever we have amplitudes directly connecting gluons of different gauge groups without fermion lines, these should vanish.
\\ \indent
Though the number of configurations tends to increase by OPP, but many of the configurations are vanishing.
Let us express an $n$-point subamplitude in (\ref{eq:total}) as,
\beq \label{eq:numberset}
\mathcal{A}_{N_1 N_2}=\sum_{k=2}^{n-2}\ub{\sum_{\text{OPP}}\sum_{\text{c.p.}} (n-k|k)}_{\text{number set}} 
\eeq
where $(n-k|k)$ denotes a configuration with $(n-k+1)$-point upper amplitude and $(k+1)$-point lower amplitude and the last sum is over possible cyclic permutations.
In (\ref{eq:numberset}), the last two sums are over all possible configurations with fixed $k$.
We call a collection of such configurations a \emph{number set}, in the sense that the number of particles in the upper and the lower amplitudes are the same in the collection.
Then we can show that in each number set there are at most two nonvanishing configurations, considering the helicity contribution of the internal line.
In result, from the first sum in (\ref{eq:numberset}), there are at most $2(n-3)$ nonvanishing configurations which contribute to an $n$-point product gauge group amplitude.
This number of configurations is equal to that of a single gauge group amplitude.
The detailed argument for this statement is given in appendix \ref{num}.
\\ \indent
On the other hand, the subamplitudes $\mathcal{A}_{N_1 N_2}$ defined in (\ref{eq:total}) are readily obtained in terms of the single gauge group multiparton amplitudes $\mathcal{A}$ involving a quark-antiquark pair \cite{mp91}, 
\beq \label{eq:singlesum}
\mathcal{A}_{N_1 N_2}(q,1,\cdots,n_1,\bar{q} ; q,\wt{1},\cdots,\wt{n}_2,\bar{q})
=\sum_{\text{OPP}}\mathcal{A}(q,1,\cdots,n_1 ; \wt{1},\cdots,\wt{n}_2,\bar{q})
\eeq
Here the sum over OPP renders all Feynman diagrams which connect directly the gluons of $SU(N_1)$ with those of $SU(N_2)$
to be canceled so that the equality is realized.
Thus in principle the amplitudes of the product gauge group can be obtained from those of single gauge group with quarks transforming as fundamental representation.
The right-hand side of (\ref{eq:singlesum}) can be calculated either using the Feynman diagrams or the recursive relations worked out in \cite{lw0501}\cite{lw0502}.
\\ \indent
However, this way of getting product gauge group amplitude is inefficient, since there are many terms in the sum of (\ref{eq:singlesum}).
In fact, for an $n$-point subamplitude, the number of configurations in the right-hand side of (\ref{eq:singlesum}) is roughly, $_{n}C_{p}$ (number of OPP) $\times$ $2(n-3)$ (number of configurations for a single gauge group amplitude), where $p$ is the number of particles of the first gauge group. 
Whereas, as mentioned above, the number of configurations in our direct recursive calculation is at most $2(n-3)$ and much smaller than that in the sum over single gauge group amplitudes in (\ref{eq:singlesum}).
\\ \indent
The final point is that in choosing the reference lines in computing the recursive relations we should choose the reference lines consistent with OPP. For example we cannot select two gluons of different gauge group as reference lines in recursive calculations. Such a selection is inconsistent with the OPP that reverses the order of any two gluons of different gauge groups, since the two reference lines cannot be permuted in the recursive relation.

\subsection{An example of configurations : $A_{34}^{+-+-}$} \label{sub:config}

\begin{figure}[t] 
\centering 
\includegraphics[scale=0.94]{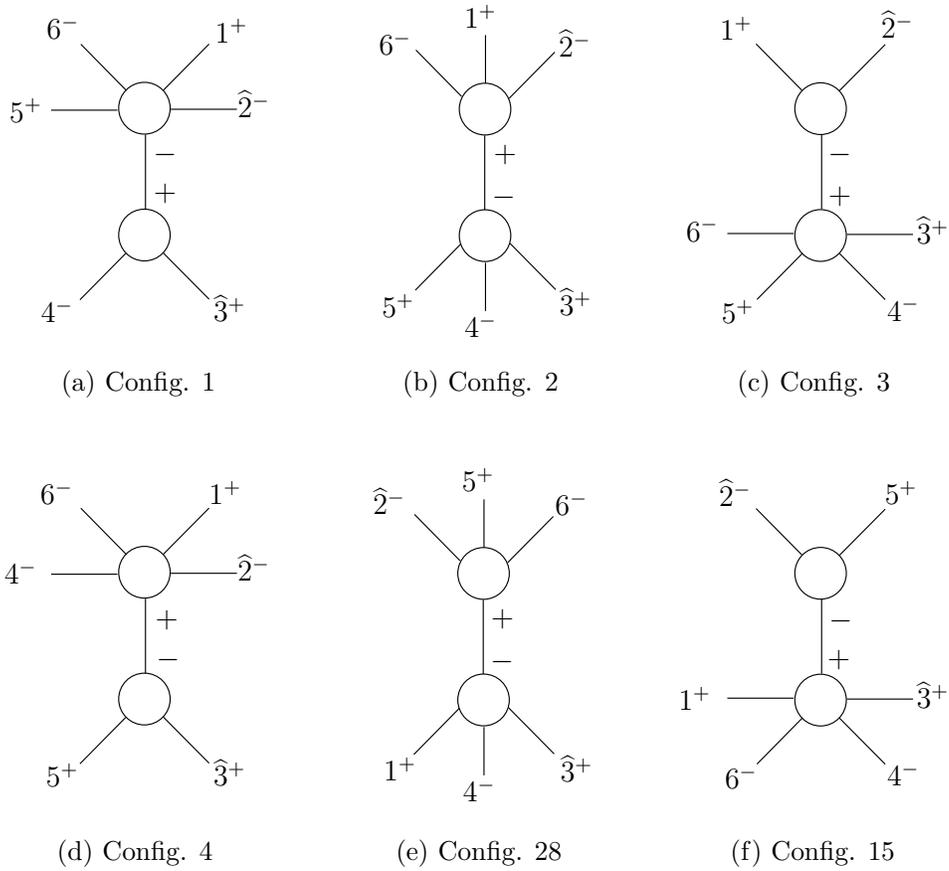}
\caption{Some configurations for $A_{34}^{+-+-}$} \label{fig:rr1}
\end{figure}

In this case, gluon $3^+$ and $4^-$ belong to the first gauge group and gluon $5^+$ and $6^-$ belong to the second.
Here the OPP generates all possible permutations of the gluons preserving the order of $3^+$ and $4^-$ and of $5^+$ and $6^-$. 
Then with these permutations and taking $2^-$(quark) and $3^+$ as reference lines, we get initially 56 configurations by the use of the of BCFW relation expressed in (\ref{eq:rr}).
All these configurations are listed in table \ref{table:config} and some of them are shown in figure \ref{fig:rr1}.
In table \ref{table:config}, each row represents the possible recursive diagrams given the order of the gluons as specified in the leftmost part and each column represents a different number set.
\begin{table}[htbp] 
\centering
\begin{tabular}{c|c|c|c} 
\hline
OPP & \multicolumn{3}{c}{configurations} \\
\hline
3456 & $(5^+6^-1^+\wh{2}^-|\wh{3}^+4^-)^1$ & $(6^-1^+\wh{2}^-|\wh{3}^+4^-5^+)^2$ & $(1^+\wh{2}^-|\wh{3}^+4^-5^+6^-)^3$ \\
\hline
3546 & $(4^-6^-1^+\wh{2}^-|\wh{3}^+5^+)^4$ & $(6^-1^+\wh{2}^-|\wh{3}^+5^+4^-)^5$ & $(1^+\wh{2}^-|\wh{3}^+5^+4^-6^-)^6$ \\
\hline
3564 & $(6^-4^-1^+\wh{2}^-|\wh{3}^+5^+)^7$ & $(4^-1^+\wh{2}^-|\wh{3}^+5^+6^-)^8$ & $(1^+\wh{2}^-|\wh{3}^+5^+6^-4^-)^9$ \\
\hline
5346 & $(4^-6^-1^+\wh{2}^-|5^+\wh{3}^+)^{10}$ & $(6^-1^+\wh{2}^-|5^+\wh{3}^+4^-)^{11}$ & $(1^+\wh{2}^-|5^+\wh{3}^+4^-6^-)^{12}$ \\                               & $(6^-1^+\wh{2}^-5^+|\wh{3}^+4^-)^{13}$ & $(1^+\wh{2}^-5^+|\wh{3}^+4^-6^-)^{14}$ & $(\wh{2}^-5^+|\wh{3}^+4^-6^-1^+)^{15}$ \\
\hline
5364 & $(6^-4^-1^+\wh{2}^-|5^+\wh{3}^+)^{16}$ & $(4^-1^+\wh{2}^-|5^+\wh{3}^+6^-)^{17}$ & $(1^+\wh{2}^-|5^+\wh{3}^+6^-4^-)^{18}$ \\
                 & $(4^-1^+\wh{2}^-5^+|\wh{3}^+6^-)^{19}$ & $(1^+\wh{2}^-5^+|\wh{3}^+6^-4^-)^{20}$ & $(\wh{2}^-5^+|\wh{3}^+6^-4^-1^+)^{21}$ \\
\hline
5634 &                                                                                                                                                          & $(4^-1^+\wh{2}^-|5^+6^-\wh{3}^+)^{22}$ & $(1^+\wh{2}^-|5^+6^-\wh{3}^+4^-)^{23}$ \\
                 & $(4^-1^+\wh{2}^-5^+|6^-\wh{3}^+)^{24}$ & $(1^+\wh{2}^-5^+|6^-\wh{3}^+4^-)^{25}$ & $(\wh{2}^-5^+|6^-\wh{3}^+4^-1^+)^{26}$ \\
                 & $(1^+\wh{2}^-5^+6^-|\wh{3}^+4^-)^{27}$ & $(\wh{2}^-5^+6^-|\wh{3}^+4^-1^+)^{28}$ & \\
\hline
\end{tabular} \caption{All possible configurations for $A_{34}^{+-+-}$} \label{table:config}
\end{table} 
\\ \indent
In the table, we numbered the configurations with superscripts and there are two configurations for each number since the internal line has two helicity compositions, $+-$ and $-+$. 
However, if either of the upper or the lower amplitude has external particles less than five, only one helicity composition of the internal line contributes.\footnote{For the 4-point amplitudes, only nonvanishing amplitudes are MHV(=$\ol{\text{MHV}}$) amplitudes. Therefore, the flip of any helicity of the nonvanishing amplitudes gives vanishing helicity compositions, such as $(-+\cdots+)$ or $(+-\cdots-)$. For the 3-point amplitudes, both $(-++)$ and $(--+)$ amplitudes are nonvanishing. However, as explained in \cite{bcf0412}, the upper amplitude of $(-++)$ type and the lower amplitude of $(--+)$ type vanish because of the choice of the reference momenta. See (\ref{eq:tool2}).}
In result, there are actually 28 configurations in this case and we are not explicit about the helicity of the internal line in the table.  
At first glance, the number of configurations is much larger (by at least 6, which is the number of OPP, times) than that of the single gauge group case with the same helicity composition. 
However, as previously mentioned, the number reduces greatly since many of them are equivalent, so counted just once, and, what is better, some of them vanish by the property that there is no interaction between the gluons of different gauge groups.
\\ \indent
Let us specify this reduction of the configurations in detail with the above example.
At first, we should find configurations generated by reducible OPP and discard them in each column of the table.
For example, let us look into the configuration 15. 
The lower amplitude of 15 is a product gauge group amplitude $A_{34}^{(1^+\wh{P}^-)3^+4^-6^-}$. 
However, this is equal to $A_{34}^{(1^+\wh{P}^-)3^+6^-4^-}$(lower amplitude of the configuration 21) and $A_{34}^{(1^+\wh{P}^-)6^-3^+4^-}$(lower amplitude of the configuration 26), since they are generated by OPP, $346\to 364$ and $346\to 634$ respectively, and an amplitude of a product gauge group theory is invariant under OPP.
That is, the configurations 21 and 26 are generated by this reducible OPP from 15 hence are equivalent.
We just keep configuration 15 for convenience.
In the similar way, we can deduce that the configurations generated by irreducible OPP are 1(=13=27), 2(=5=11), 3(=6=9=12=18=23), 4(=7=10=16), 8(=17=22), 14(=20=25), 15(=21=26), 19(=24) and 28. See table \ref{table:irrconfig}.
\begin{table}[tbp] 
\centering
\begin{tabular}{c|c|c|c} 
\hline
OPP & \multicolumn{3}{c}{configurations} \\
\hline
3456 & $(5^+6^-1^+\wh{2}^-|\wh{3}^+4^-)^1$ & $(6^-1^+\wh{2}^-|\wh{3}^+4^-5^+)^2$ & $(1^+\wh{2}^-|\wh{3}^+4^-5^+6^-)^3$ \\
\hline
3546 & $(4^-6^-1^+\wh{2}^-|\wh{3}^+5^+)^4$ & &  \\
\hline
3564 & & $(4^-1^+\wh{2}^-|\wh{3}^+5^+6^-)^8$ &  \\
\hline
5346 & & $(1^+\wh{2}^-5^+|\wh{3}^+4^-6^-)^{14}$ & $(\wh{2}^-5^+|\wh{3}^+4^-6^-1^+)^{15}$ \\
\hline
5364 & $(4^-1^+\wh{2}^-5^+|\wh{3}^+6^-)^{19}$ & &  \\
\hline
5634 & & $(\wh{2}^-5^+6^-|\wh{3}^+4^-1^+)^{28}$ & \\
\hline
\end{tabular} \caption{Configurations related by irreducible OPP for $A_{34}^{+-+-}$. Most of them vanish except 1 and 28.} \label{table:irrconfig}
\end{table} 
\\ \indent
Moreover, if either of the upper or the lower amplitude of one configuration has gluons of different gauge groups without any fermion($1^+$ or $2^-$), this configuration vanishes.
This originates from the fact that there is no gluonic vertex where the gluons of different gauge groups are connected in the Feynman diagrams of the product gauge group theory.
For instance, configuration 4 vanishes since there is no vertex connecting gluon 4 and 5, which belong to different gauge groups. 
Configurations 2, 3, 8, 14, and 19 are all discarded for the similar reason.
\\ \indent
Now, there remain only 3 configurations, 1, 15 and 28. 
However, configuration 15 vanishes since the helicity along the fermion line should be conserved. 
In detail, the upper amplitude of 15 is $A_S^{(\wh{P}^-\wh{2}^-)5^+}$, whereas $\wh{P}$ should be an antiquark and has positive helicity to have a nonvanishing amplitude.
\\ \indent
Finally, there remain just two nonvanishing configurations, 1 and 28.
As can be seen in table \ref{table:irrconfig}, these are the only nonvanishing configurations in each column, i.e. in each number set.
This coincides with the statement mentioned above that there are at most two (allowing the two helicity compositions of the internal line) nonvanishing configurations in each number set.
The general argument for the number of configurations appears in appendix \ref{num}.

\subsection{A Calculation Example : $A_{34}^{+-+-}$}

Once we construct all configurations for the recursive relation, the rest calculation is just the same as the gluonic and the single gauge group cases. 
In a configuration, the upper and the lower amplitudes have the particles of shifted momenta $\wh{p}_m$, $\wh{p}_n$ and $\wh{P}$. 
Then, in the sense that the final form of the amplitude should be expressed with the spinors of the external momenta only, we need to express the spinors of such shifted momenta as \cite{bcf0412}, 
\begin{align} \label{eq:tool2}
%\begin{split}
& \lambda_{\wh{m}} = \lambda_{m}, \qquad 
                \wt{\lambda}_{\wh{m}} = \wt{\lambda}_{m} - \frac{P^2}{\la m|P|n]} \wt{\lambda}_{n}, \nn \\
& \lambda_{\wh{n}} = \lambda_{n} + \frac{P^2}{\la m|P|n]} \lambda_{m}, \qquad 
                \wt{\lambda}_{\wh{n}} = \wt{\lambda}_{n}, \\
& \la \bu \wh{P} \ra = \frac{\la \bu |{P}|n]}{[\wh{P}n]}, \qquad
                [\wh{P} \bu] = \frac{\la m|P|\bu]}{\la m\wh{P}\ra} \nn
%\end{split}
\end{align}
The denominators in the last line, $[\wh{P}n]$ and $\la m\wh{P}\ra$, appear always in pair in each configuration as $\la m\wh{P}\ra[\wh{P}n] = \la m|P|n]$, since each configuration should be invariant under the rescaling of the spinors of $\wh{P}$, $\lambda_{\wh{P}} \to t \lambda_{\wh{P}}$ and $\wt{\lambda}_{\wh{P}} \to t^{-1} \wt{\lambda}_{\wh{P}}$.
\\ \indent
Let us calculate the configurations of the example of the previous subsection, $A_{34}^{+-+-}$. 
There were two nonvanishing configurations, $(5^+6^-1^+\wh{2}^-|\wh{3}^+4^-)$ and $(\wh{2}^-5^+6^-|\wh{3}^+4^-1^+)$.
The first one is,
\begin{align}
%\begin{split}
(5^+6^-1^+\wh{2}^-|\wh{3}^+4^-) 
&= A_{\wh{P}}^{(1^+\wh{2}^-)\wh{P}^-5^+6^-} \frac{1}{t_{34}} A_G^{-\wh{P}^+\wh{3}^+4^-}  \nn \\
&= \frac{-[15]^3[\wh{2}5]}{[\wh{2}\wh{P}][\wh{P}1][\wh{2}5][56][61]}
                \frac{1}{\la 34\ra[34]}      
                \frac{[\wh{P}\wh{3}]^3}{[\wh{3}4][4\wh{P}]}  \\
&= \frac{[15]^3\la 24\ra^3}{t_{234}\la 23\ra\la 34\ra[56][61]\la 2|3+4|1]} \nn
%\end{split}
\end{align}
where the last equality was obtained applying (\ref{eq:tool2}) to this case as,
\begin{align}
%\begin{split}
\wt{\lambda}_{\wh{2}} &= \wt{\lambda}_2 - \frac{t_{34}}{\la 2|3+4|3]}\wt{\lambda}_3,  \nn \\
\wt{\lambda}_{\wh{3}} &= \wt{\lambda}_3, \\
\quad [\wh{P} \bu] &= \frac{\la 2|3+4|\bu]}{\la 2\wh{P}\ra} \nn
%\end{split}
\end{align}
Similarly, we get the second configuration,
\begin{align}
%\begin{split}
(\wh{2}^-5^+6^-|\wh{3}^+4^-1^+)
&= A_S^{(\wh{P}^+\wh{2}^-)5^+6^-} \frac{1}{t_{341}} A_S^{(1^+-\wh{P}^-)\wh{3}^+4^-}  \nn \\
&= \frac{\la \wh{2}6\ra^3\la \wh{P}6\ra}{\la \wh{2}5\ra\la 56\ra\la 6\wh{P}\ra\la \wh{P}\wh{2}\ra}
                \frac{1}{t_{341}} 
                \frac{[1\wh{3}]^3[\wh{P}\wh{3}]}{[\wh{P}\wh{3}][\wh{3}4][41][1\wh{P}]} \\
&= \frac{\la 26\ra^3[13]^3}{t_{341}\la 25\ra\la 56\ra[34][41]\la 2|3+4|1]} \nn
%\end{split}
\end{align}
where we used
\begin{align}
%\begin{split}
\lambda_{\wh{2}} &= \lambda_2, \nn \\
\wt{\lambda}_{\wh{3}} &= \wt{\lambda}_3, \nn \\
\la \bu \wh{P}\ra &= \frac{\la \bu|3+4+1|3]}{[\wh{P}3]}, \\
\quad [\wh{P} \bu] &= \frac{\la 2|3+4+1|\bu]}{\la 2\wh{P}\ra} \nn 
%\end{split}
\end{align}
\indent
The above result for $A_{34}^{+-+-}$ agrees with the result by the direct use of Feynman rules and with the sum over corresponding single gauge group amplitudes as in (\ref{eq:singlesum}).

\subsection{Getting Amplitudes from Other Amplitudes : Exchanging Gluon Indices} \label{exch}

As can be seen in (\ref{eq:total}) and (\ref{eq:singlesum}), an amplitude of the product gauge group theory is invariant under OPP.
For example, $A_{34}^{3^-5^+4^-6^+}$ is equal to $A_{34}^{3^-4^-5^+6^+}$. 
Using this property of product gauge group amplitudes, we can calculate all amplitudes from some \emph{primitive amplitudes}, which are defined below, just by exchanging gluon indices appropriately. 
For instance, we deduce that $A_{35}^{3^-4^+5^-6^+}$ can be obtained from $A_{34}^{3^-4^-5^+6^+}$ by exchanging gluon indices 4 and 5, since $A_{35}^{3^-4^+5^-6^+}$ is equal to $A_{35}^{3^-5^-4^+6^+}$, which becomes  $A_{34}^{3^-4^-5^+6^+}$ with 4 and 5 exchanged. 
In the similar way, we easily notice that $A_{345}^{3^-4^-5^+6^+}$ is equal to $A_{3}^{3^+5^-4^-6^+}$ with $3 \to 6$, $4 \to 3$, $5 \to 4$ and $6 \to 5$.
\\ \indent
Then we can define, by convention, the primitive amplitudes as the amplitudes in which gluons of different gauge groups are completely split and the number of the the first gauge group gluons is less than or equal to the half of the total number of the gluons. 
In the case of 4-point and 5-point amplitudes, there is just one class of primitive amplitudes, $A_{3}^{\cdots}$.
For 6-point and 7-point cases, there are two, $A_{3}^{\cdots}$ and $A_{34}^{\cdots}$. 
In general, for N-point function, there are $[N/2-1]$ classes of primitive amplitudes. 
In this paper, we only give the results of the primitives.
\\ \indent
We calculated all primitive MHV amplitudes and six and seven point primitive next-to-MHV amplitudes and compared the results with the sums of single gauge group amplitudes numerically, by the use of (\ref{eq:singlesum}). 
For the sum over the single gauge group amplitudes, we used previous results of 6-point amplitudes in \cite{lw0501} and new results of 7-point amplitudes (appendix~\ref{7ps}) which were calculated recursively as well. 
All results are given in the section~\ref{mhv}, \ref{6p} and appendix \ref{7p}.

\section{All MHV Amplitudes} \label{mhv}

For the convenience of the proof, all primitive MHV amplitudes can be divided into several cases according to the number of the gluons that belong to each gauge group and the location of the $-$ helicity gluon. 
In table~\ref{table:mhv} are listed these cases with reference momenta and the corresponding configurations.
In the table, $a$ is the gluon with $-$ helicity and $i$, $j$ are the numbers of the gluons of the first and second gauge group respectively. 
For example, in case 5, gluon 3 belongs to the first gauge group and gluon 4, 5, $\cdots$ belong to the second, where the $-$ helicity gluon is one of 4, 5, $\cdots$. 
As before, particle $1^+$ and $2^-$ are the antiquark and the quark respectively. 
Similar conventions are used in the other cases.

\begin{table}[htbp] 
\centering
\begin{tabular}{c|c|c|c}
\hline  & case & ref. line & configuration \\
\hline
 1 &  $i \ge 2$, $a \ge 5$ & 2, 3 & $(\cdots a^-\cdots 1^+\wh{2}^-|^{+-}|\wh{3}^+4^+)$ \\
 2 &  $i \ge 2$, $a=4$ & 2, 3 & $(\cdots 1^+\wh{2}^-|^{-+}|\wh{3}^+4^-)$ \\
 3 &  $i \ge 3$, $a=3$ & 3, 4 & $(\cdots 1^+2^-\wh{3}^-|^{+-}|\wh{4}^+5^+)$ \\
 4 &  $i=2$, $a=3$ & 3, 4 & $(2^-\wh{3}^-\cdots|^{+-}|\wh{4}^+1^+)$ \\
 5 &  $i=1$, $j \ge 2$, $a \ge 4$ & 2, 3 & $(\wh{2}^-\cdots a^-\cdots|^{+-}|\wh{3}^+1^+)$ \\
 6 &  $i=1$, $j \ge 2$, $a=3$ & 2, 4 & $(\cdots\wh{2}^-3^-|^{+-}|\wh{4}^+5^+)$ \\
 7 &  $i=j=1$, $a=4$ & 2, 3 & $(\wh{2}^-4^-|^{+-}|\wh{3}^+1^+)$ \\
 8 &  $i=j=1$, $a=3$ & 2, 4 & $(\wh{2}^-3^-|^{+-}|\wh{4}^+1^+)$ \\
\hline
\end{tabular} \caption{Configurations for all MHV amplitudes} \label{table:mhv}
\end{table}

\indent
In each case there is just one configuration by judicious choice of the reference lines, giving expected simple result for the MHV amplitude,

\beq \label{eq:mhv}
-\frac{\la 2a\ra^3\la 1a\ra}{\ub{\la 2\bu\ra \cdots \la\bu 1\ra}_{SU(N_1)}\ub{\la 2\bu\ra \cdots \la\bu 1\ra}_{SU(N_2)}}
\eeq 
where $\bu$ represent the gluons of each gauge group in the order specified by the color factorization.
For the $\ol{\textrm{MHV}}$ amplitudes, the similar construction of configurations gives expected result,
\beq \label{eq:mhvb}
\frac{[1a]^3[2a]}{\ub{[2\bu]\cdots[\bu 1]}_{SU(N_1)}\ub{[2\bu]\cdots[\bu 1]}_{SU(N_2)}}
\eeq 
In the followings, we provide the calculation results of the configurations in table~\ref{table:mhv}.
Each result has the form of (\ref{eq:mhv}).
\begin{align}
%\begin{split}
\text{case 1}
&= \frac{-\la 2a\ra^3\la 1a\ra}{\la 2\wh{P}\ra\la \wh{P}\bu\ra\cdots\la \bu 1\ra\la 2\bu\ra\cdots\la \bu 1\ra}
    \frac{1}{\la 34\ra[34]} \frac{[34]^3}{[4\wh{P}][\wh{P}3]} \nn \\
&= -\frac{\la 2a\ra^3\la 1a\ra}{\la 23\ra\la 34\ra\la 4\bu\ra\cdots\la \bu 1\ra\la 2\bu\ra\cdots\la \bu 1\ra}
%\end{split} 
\\ \nn \\
%\begin{split}
\text{case 2}
&= \frac{-\la 2\wh{P}\ra^3\la 1\wh{P}\ra}{\la 2\wh{P}\ra\la \wh{P}\bu\ra\cdots\la \bu 1\ra\la 2\bu\ra\cdots\la \bu 1\ra}
    \frac{1}{\la 34\ra[34]} \frac{[\wh{P}3]^3}{4\wh{P}][34]} \nn \\
&= -\frac{\la 24\ra^3\la 14\ra}{\la 23\ra\la 34\ra\la 4\bu\ra\cdots\la \bu 1\ra\la 2\bu\ra\cdots\la \bu 1\ra}
%\end{split}
\\ \nn \\
%\begin{split} 
\text{case 3}
&= \frac{-\la 23\ra^3\la 13\ra}{\la 23\ra\la 3\wh{P}\ra\la \wh{P}\bu\ra\cdots\la \bu 1\ra\la 2\bu\ra\cdots\la \bu 1\ra}
    \frac{1}{\la 45\ra[45]} \frac{[45]^3}{[5\wh{P}][\wh{P}4]} \nn \\
&= -\frac{\la 23\ra^3\la 13\ra}{\la 23\ra\la 34\ra\la 4\bu\ra\cdots\la \bu 1\ra\la 2\bu\ra\cdots\la \bu 1\ra}
%\end{split}
\\ \nn \\
%\begin{split}
\text{case 4}
&= \frac{-\la 23\ra^3\la \wh{P}3\ra}{\la 23\ra\la 3\wh{P}\ra\la 2\bu\ra\cdots\la \bu\wh{P}\ra}
    \frac{1}{\la 41\ra[41]} \frac{[14]^2[\wh{P}4]}{[\wh{P}4][1\wh{P}]} \nn \\
&= -\frac{\la 23\ra^3\la 13\ra}{\la 23\ra\la 34\ra\la 41\ra\la 2\bu\ra\cdots\la \bu 1\ra}
%\end{split}
\\ \nn \\
%\begin{split}
\text{case 5}
&= \frac{\la 2a\ra^3\la \wh{P}a\ra}{\la 2\bu\ra\cdots\la \bu\wh{P}\ra\la \wh{p}2\ra}
    \frac{1}{\la 31\ra[31]} \frac{[13]^3[\wh{P}3]}{[\wh{P}3][31][1\wh{P}]} \nn \\
&= -\frac{\la 2a\ra^3\la 1a}{\la 23\ra\la 31\ra\la 2\bu\ra\cdots\la \bu 1\ra}
%\end{split} 
\\ \nn \\
%\begin{split}
\text{case 6}
&= \frac{-\la 23\ra^3\la 13\ra}{\la 23\ra\la 31\ra\la 2\wh{P}\ra\la \wh{P}\bu\ra\cdots\la \bu 1\ra}
    \frac{1}{\la 45\ra[45]} \frac{[45]^3}{[5\wh{P}][\wh{P}4]} \nn \\
&= -\frac{\la 23\ra^3\la 13\ra}{\la 23\ra\la 31\ra\la 24\ra\la 4\bu\ra\cdots\la \bu 1\ra}
%\end{split}
\\ \nn \\
%\begin{split}
\text{case 7}
&= \frac{\la 24\ra^3\la \wh{P}4\ra}{\la 24\ra\la 4\wh{P}\ra\la \wh{P}2\ra}
    \frac{1}{\la 31\ra[31]} \frac{[13]^3[\wh{P}3]}{\wh{P}3][31][1\wh{P}]} \nn \\
&= -\frac{\la 24\ra^3\la 14\ra}{\la 23\ra\la 31\ra\la 24\ra\la 41\ra}
%\end{split} 
\\ \nn \\
%\begin{split}
\text{case 8}
&= (3\leftrightarrow 4)\text{ of case7} \nn \\
&= -\frac{\la 23\ra^3\la 13\ra}{\la 23\ra\la 31\ra\la 24\ra\la 41\ra} 
%\end{split} 
\end{align}

\section{6-Point Next-to-MHV Amplitudes} \label{6p}

In this section, we present the configurations and the calculation results of 2 primitive classes of 6-point next-to-MHV amplitudes, $A_{34}$ and $A_{3}$. For each amplitude, we provide the corresponding nonvanishing configurations and the calculation result.
All results are compared to the sum over the corresponding single gauge group amplitudes via the equality described in (\ref{eq:singlesum}).
We give the examples of this equality for the case of $A_{34}^{3^+4^-5^+6^-}$ and $A_{3}^{3^+4^+5^-6^-}$ in table \ref{table:opp}. 
In each column of the table, the product gauge group amplitude in the first row is the sum over the single gauge group amplitudes in the rows below.
For other product gauge group amplitudes, we can similarly find the corresponding single gauge group amplitudes which contribute to the sum.
\begin{table}[htbp]
\centering
\begin{tabular}{c|c|c}
\hline  & $A_{34}^{3^+4^-5^+6^-}$ & $A_{3}^{3^+4^+5^-6^-}$ \\
\hline $ $ & $A_S^{3^+4^-5^+6^-}$ & $A_S^{3^+4^+5^-6^-}$ \\
     $ $ & $A_S^{5^+6^-3^+4^-}$ & $A_S^{4^+3^+5^-6^-}$ \\
$\sum A_S$ & $A_S^{3^+5^+4^-6^-}$ & $A_S^{4^+5^-3^+6^-}$ \\
     $ $ & $A_S^{3^+5^+6^-4^-}$ & $A_S^{4^+5^-6^-3^+}$ \\
     $ $ & $A_S^{5^+3^+4^-6^-}$ &  \\
     $ $ & $A_S^{5^+3^+6^-4^-}$ &  \\
\hline
\end{tabular} \caption{Product gauge group amplitude as a sum over single gauge group amplitudes} \label{table:opp}
\end{table}
\\ \indent
In the case of $A_{34}$, two of the amplitudes are obtained from others by using the scheme of section~\ref{exch}. 
For instance, $A_{34}^{3^-4^-5^+6^+}$ is obtained from $A_{34}^{3^+4^+5^-6^-}$ by exchanging 3 with 5 and 4 with 6,
since $A_{34}^{3^+4^+5^-6^-}$ is equal to $A_{34}^{5^-6^-3^+4^+}$. 
Similarly, we get $A_{34}^{3^-4^+5^+6^-}$ from $A_{34}^{3^+4^-5^-6^+}$ by the same substitution.
In the case of $A_{3}$, there are no relationships like these. 
We can get any other amplitude from these 2 classes of primitive amplitudes. 
In each helicity composition, the configurations are obtained using the similar computations of section \ref{rr}.
As shown below, there are only 2 or 3 configurations for each amplitude.

\subsection{$A_{34}$}

\begin{align}
%\begin{split}
A_{34}^{--++}
&= (3^-4^-1^+\wh{2}^-|^{+-}|\wh{5}^+6^+)
                +(\wh{2}^-3^-|^{+-}|\wh{5}^+6^+4^-1^+) \nn \\
&= \frac{\la 2|3+4|1]^2}{\la 25\ra\la 56\ra[34][41]\la 6|2+5|3]}
                +\frac{\la 4|2+3|5]^2}{t_{235}\la 61\ra[23]\la 6|2+5|3]}
%\end{split} 
\\ \nn \\ 
%\begin{split}
A_{34}^{-+-+}
&= (2^-\wh{3}^-|^{+-}|\wh{4}^+5^-6^+1^+)
                +(2^-\wh{3}^-5^-6^+|^{+-}|\wh{4}^+1^+) \nn \\
&= -\frac{\la 51\ra[24]\la 5|2+3|4]^2}{t_{234}[23][34]\la 56\ra\la 61\ra\la 1|3+4|2]}
    -\frac{\la 31\ra[26]\la 3|1+4|6]^2}{t_{134}[25][56]\la 34\ra\la 41\ra\la 1|3+4|2]}
%\end{split} 
\\ \nn \\ 
%\begin{split}
A_{34}^{-++-}
&= (3^-4^+1^+\wh{2}^-|^{-+}|\wh{5}^+6^-)
    +(\wh{2}^-3^-4^+|^{+-}|\wh{5}^+6^-1^+)
    +(\wh{2}^-3^-|^{+-}|\wh{5}^+6^-4^+1^+) \nn \\
&= -\frac{[14]^2\la 26\ra^3\la 6|2+5|4]}{t_{256}\la 25\ra\la 56\ra[34]\la 2|5+6|1]\la 6|2+5|3]} \nn \\
&       \quad +\frac{\la 23\ra^3[15]^3\la 3|6+1|5]}{t_{156}[56][61]\la 34\ra\la 4|6+1|5]\la 2|5+6|1]} \nn \\
& \quad +\frac{\la 6|2+3|5]^3}{t_{235}[23]\la 41\ra\la 6|2+5|3]\la 4|2+3|5]}
%\end{split} 
\\ \nn \\ 
%\begin{split}
A_{34}^{+--+}
&= A_{34}^{-++-}(3\leftrightarrow 5,4\leftrightarrow 6) 
%\end{split} 
\\ \nn \\ 
%\begin{split}
A_{34}^{+-+-} 
&= (5^+6^-1^+\wh{2}^-|^{-+}|\wh{3}^+4^-)
                +(\wh{2}^-5^+6^-|^{+-}|\wh{3}^+4^-1^+) \nn \\
&= \frac{\la 24\ra^3[15]^3}{t_{234}\la 23\ra\la 34\ra[56][61]\la 2|3+4|1]}
    -\frac{\la 26\ra^3[13]^3}{t_{341}\la 25\ra\la 56\ra[34][41]\la 2|3+4|1]}
%\end{split} 
\\ \nn \\ 
%\begin{split}
A_{34}^{++--}
&= A_{34}^{--++}(3\leftrightarrow 5,4\leftrightarrow 6) 
%\end{split} 
\end{align}

\subsection{$A_{3}$}

\begin{align}
%\begin{split}
A_{3}^{--++}
&= (1^+2^-3^-\wh{4}^-|^{+-}|\wh{5}^+6^+)
                +(2^-\wh{4}^-|^{+-}|3^-\wh{5}^+6^+1^+) \nn \\
&= \frac{\la 4|5+6|2]\la 4|5+6|1]^2}{t_{123}[23][31]\la 45\ra\la 56\ra\la 6|5+4|2]}
                +\frac{[25]\la 3|2+4|5]^2}{t_{245}[24][45]\la 61\ra\la 6|5+4|2]}
%\end{split} 
\\ \nn \\ 
%\begin{split}
A_{3}^{-+-+}
&= (2^-3^-4^+\wh{5}^-|^{+-}|\wh{6}^+1^+)
                +(4^+\wh{5}^-|^{-+}|\wh{6}^+1^+2^-3^-)
                +(2^-4^+\wh{5}^-|^{+-}|3^-\wh{6}^+1^+) \nn \\
&= -\frac{\la 51\ra\la 5|6+1|4]^3}{t_{234}[23]\la 56\ra\la 61\ra\la 5|6+1|3]\la 1|6+5|4]} \nn \\
&       \quad +\frac{\la 23\ra^2[46]^4}{t_{123}[45][56]\la 2|4+5|6]\la 1|6+5|4]} \nn \\
& \quad +\frac{[16]^2\la 25\ra^3\la 5|3+1|6]}{t_{245}\la 24\ra\la 45\ra[31]\la 2|3+1|6]\la 5|6+1|3]}
%\end{split} 
\\ \nn \\ 
%\begin{split}
A_{3}^{-++-}
&= (6^-1^+\wh{2}^-3^-|^{+-}|\wh{4}^+5^+) 
                +(1^+\wh{2}^-3^-|^{+-}|\wh{4}^+5^+6^-)
                +(\wh{2}^-3^-|^{+-}|\wh{4}^+5^+6^-1^+) \nn \\
&= \frac{\la 23\ra^2[45]^3}{t_{123}[56]\la 1|5+6|4]\la 2|4+5|6]} \nn \\
&       \quad -\frac{\la 2|4+5|1]^3}{\la 24\ra\la 45\ra[61][13]\la 5|2+4|3]\la 2|4+5|6]} \nn \\
&       \quad +\frac{\la 6|2+3|4]^3}{t_{234}[23]\la 56\ra\la 5|2+4|3]\la 1|2+3|4]}
%\end{split} 
\\ \nn \\ 
%\begin{split}
A_{3}^{+--+}
&= (2^-3^+4^-\wh{5}^-|^{+-}|\wh{6}^+1^+)
                +(4^-\wh{5}^-|^{+-}|\wh{6}^+1^+2^-3^+) \nn \\
&= -\frac{\la 51\ra\la 5|6+1|3]^2}{t_{561}\la 56\ra\la 61\ra[24]\la 1|6+5|4]}
                -\frac{\la 1|4+5|6]\la 2|4+5|6]^2}{t_{123}\la 23\ra\la 31\ra[45][56]\la 1|6+5|4]}
%\end{split} 
\\ \nn \\ 
%\begin{split}
A_{3}^{+-+-}
&= (1^+2^-3^+\wh{4}^-|^{-+}|\wh{5}^+6^-)
                +(2^-3^+\wh{4}^-|^{+-}|\wh{5}^+6^-1^+)
                +(2^-\wh{4}^-|^{+-}|3^+\wh{5}^+6^-1^+) \nn \\
&= -\frac{[13]^2\la 46\ra^4}{t_{123}\la 45\ra\la 56\ra\la 6|4+5|2]\la 4|5+6|1]} \nn \\
&       \quad -\frac{\la 24\ra^2[15]^3\la 4|6+1|5]}{t_{234}\la 23\ra[56][61]\la 3|6+1|5]\la 4|5+6|1]} \nn \\
& \quad +\frac{[25]\la 6|2+4|5]^3}{t_{245}[24][45]\la 31\ra\la 3|2+4|5]\la 6|4+5|2]}
%\end{split}
\\ \nn \\ 
%\begin{split}
A_{3}^{++--}
&= (1^+\wh{2}^-3^+|^{-+}|\wh{4}^+5^-6^-)
                +(6^-1^+\wh{2}^-3^+|^{-+}|\wh{4}^+5^-) \nn \\
&= -\frac{\la 2|5+6|4]^3}{t_{123}\la 23\ra\la 31\ra[45][56]\la 2|4+5|6]}
                -\frac{[13]^2\la 25\ra^3}{t_{245}\la 24\ra\la 45\ra[61]\la 2|4+5|6]|}
%\end{split}
\end{align}

\section{CSW Relations} \label{csw}

\begin{figure}[htbp]
\centering
\includegraphics[scale=0.94]{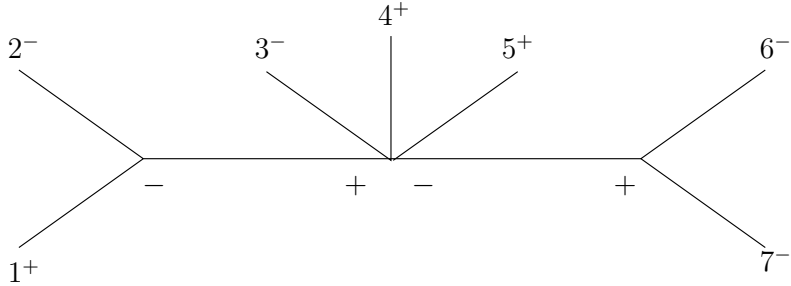}
\caption{CSW diagram of $(1^+2^-)(3^-4^+5^+)(6^-7^-)$} \label{fig:csw}
\end{figure}

We denote the diagrams of the CSW calculation as, 
\beq
\ub{(\cdots i^{\pm} \cdots)}_{\text{MHV vertex}}(\cdots j^{\pm}\cdots)\cdots (\cdots)
\eeq
where each part enclosed by parentheses represents a certain MHV vertex and $i^{\pm}$, $j^{\pm}$, $\cdots$ denote the external particles in each MHV vertex.
For example, $(1^+2^-)(3^-4^+5^+)(6^-7^-)$ represents the diagram in fig~\ref{fig:csw}.
\\ \indent
The CSW diagrams for the product gauge group amplitude are obtained in similar way that the configurations of the recursive relations are constructed.
That is, we weave all possible MHV vertices considering all possible OPP, that is, allowing all permutations between gluons of different gauge group, but preserving the order of gluons of each gauge group. 
In doing this, however, the diagrams that have any vertex where different gauge group gluons are connected without quark or antiquark are discarded.
In addition, some diagrams any of whose vertices includes quark and antiquark of same helicity are also discarded, as in the recursive calculations.
Some examples of the CSW diagrams and their calculation results are given below. 
In the expression, $w$ denotes an arbitrary antiholomorphic spinor.
\begin{align}
%\begin{split}
A_{3}^{--+}
&= (2^-3^-)(4^-5^+1^+)+(4^-5^+)(1^+2^-3^-)+(2^-4^-)(3^-5^+1^+)+(3^-1^+)(2^-4^-5^+) \nn \\
&= \frac{\la 4|2+3|w]^2\la 14\ra}{[3w][23]\la 45\ra\la 51\ra\la 1|2+3|w]} \nn \\
& \quad  -\frac{[5w]^3\la 23\ra^2}{[4w][45]\la 2|1+3|w]\la 1|2+3|w]} \nn \\
& \quad  -\frac{\la 3|2+4|w]^2}{[4w][24]\la 51\ra\la 5|2+4|w]} \nn \\
& \quad  -\frac{[1w]^2\la 24\ra^2\la 4|1+3|w]}{[3w][13]\la 45\ra\la 2|1+3|w]\la 5|1+3|w]} \nn \\
&= \frac{[15]^3[25]}{[23][31][24][45][51]} 
%\end{split}
\end{align}
\indent
In the above, for example, the first diagram $(2^-3^-)(4^-5^+1^+)$ is calculated as,
\begin{align}
\frac{\la 23\ra^3 \la P3\ra}{\la 23\ra\la 3P\ra\la P2\ra}
\frac{1}{\la 23\ra^3[23]}
\frac{\la P4\ra^3\la 14\ra}{\la P4\ra\la 45\ra\la 51\ra\la 1P\ra}
&= \frac{\la 4|2+3|w]^2\la 14\ra}{[3w][23]\la 45\ra\la 51\ra\la 1|2+3|w]}
\end{align}
where $|P \ra = (p_2+p_3)\cdot w = |2\ra[2w]+|3\ra[3w]$ was used.
If we take $w$ as one of $|i]$, we get the final equality (numerically), which is a correct $\ol{\text{MHV}}$ amplitude for the product gauge group theory as expressed in (\ref{eq:mhvb}). 
We give below the calculation results of two six point next-to-MHV amplitudes with different helicity compositions.
Each amplitude agrees with the result calculated by the use of recursive relations in section \ref{6p}.
\begin{align}
%\begin{split}
A_{34}^{--++}
&= (3^-4^-)(5^+6^+1^+2^-)+(2^-3^-)(4^-5^+6^+1^+) \nn \\
& \quad +(4^-6^+1^+)(2^-3^-5^+)+(4^-1^+)(2^-3^-5^+6^+) \nn \\
&= -\frac{\la 2|3+4|w]^2}{[3w][4w][34][25]\la 56\ra\la 61\ra} \nn \\
& \quad -\frac{\la 4|2+3|w]^2}{[3w][23]\la 56]\la 61\ra\la 5|2+3|w]} \nn \\
& \quad +\frac{\la 23\ra^2\la 4|1+6|w]^2}{t_{235}\la 61]\la 25]\la 6|1+4w]\la 5|2+3|w]} \nn \\
& \quad +\frac{[1w]^2\la 23\ra^2}{[4w][41]\la 25\ra\la 56\ra\la 6|1+4|w]} \nn \\
&= \frac{\la 2|3+4|1]^2}{\la 25\ra\la 56\ra[34][41]\la 6|2+5|3]}
                +\frac{\la 4|2+3|5]^2}{t_{235}\la 61\ra[23]\la 6|2+5|3]}
%\end{split} 
\\ \nn \\
%\begin{split}
A_{34}^{-+-+}
&= (5^-6^+)(1^+2^-3^-4^+)+(3^-4^+)(5^-6^+1^+2^-)+(2^-3^-)(4^+5^-6^+1^+) \nn \\
& \quad +(2^-5^+)(3^-4^+6^+1^+)+(5^-6^+1^+)(2^-3^-4^+)+(3^-4^+1^+)(2^-5^-6^+) \nn \\
&= \frac{[6w]^3\la 23\ra^2\la 13\ra}{[5w][56]\la 34\ra\la 41\ra\la 2|5+6|w]\la 1|5+6|w]} \nn \\
& \quad +\frac{\la 15\ra\la 5|2+3|w]^2}{[3w][23]\la 41\ra\la 56\ra\la 61\ra\la 4|2+3|w]} \nn \\
& \quad -\frac{\la 15\ra\la 23\ra^2\la 5|6+1|w]^2\la 3|2+4|w]}{t_{234}\la 56\ra\la 61\ra\la 34\ra\la 1|5+6|w]\la 2|3+4|w]\la 4|2+3|w]} \nn \\
& \quad +3\leftrightarrow 5,4\leftrightarrow 6 \text{ of above terms} \nn \\
&= -\frac{\la 51\ra[24]\la 5|2+3|4]^2}{t_{234}[23][34]\la 56\ra\la 61\ra\la 1|3+4|2]}
    -\frac{\la 31\ra[26]\la 3|1+4|6]^2}{t_{134}[25][56]\la 34\ra\la 41\ra\la 1|3+4|2]} \\ \nn
%\end{split}
\end{align}

\section*{Acknowledgments}
This work is supported by the Korea Science and Engineering Foundation(KOSEF) Grant R01-2004-000-10526-0 (JP, WS) and by the Science Research Center Program of KOSEF through the Center for Quantum Spacetime(CQUeST) of Sogang University with grant number R11-2005-021 (JP).

\section*{Appendix}
\appendix

\section{Number of Configurations} \label{num}

In general, if we put aside the helicity effects which make some configurations vanishing, the number of configurations in a product gauge group amplitude does not exceed that of a single gauge group amplitude.
This can be shown in the following way. 
\\ \indent
In section \ref{rr} we constructed a collection of all configurations with the same number of particles in the upper(or lower) amplitude and denoted the collection as a \emph{number set}.  
For a single gauge group amplitude, if we take two neighboring particles as reference lines, there are evidently at most two (considering the two helicity compositions, $(-+)$ and $(+-)$, of the internal line) configurations in a number set. 
Then how many configurations are there in a number set of a product gauge group amplitude? 
To check this, it is convenient to classify the configurations into two cases according to the helicity of the first gluon (namely $3$) of either gauge group. 
\\ \indent
In the first case, where $3$ has positive helicity, let us take $2^-$ and $3^+$ as reference lines. 
Then there are, according to the locations of $1^+$(antiquark) and $2^-$(quark), possibly two types of configurations,
\begin{align} \label{eq:case1}
\begin{split}
(\widehat{2}^-\ub{\cdots}_{g'} | \ub{\widehat{3}^+\cdots}_{g}\ub{\cdots}_{g'} 1^+) \\
(\ub{\cdots}_{g} 1^+\widehat{2}^-\ub{\cdots}_{g'} | \ub{\widehat{3}^+\cdots}_{g}) 
\end{split}
\end{align}
where $g$ denote the gluons of the gauge group to which $3^+$ belongs and $g'$ the gluons of the other gauge group.
Note that two types of configurations in (\ref{eq:case1}) belong to different number sets.
\\ \indent
The second case, where $3$ has negative helicity, has at least one pair of adjacent gluons, $n^-$ and $(n+1)^+$, which are in the same gauge group. If we take this two gluons as reference particles, there are three types of configurations,
\begin{align} \label{eq:case2}
\begin{split}
(\ub{\cdots\widehat{n}^-}_{g} | \ub{\widehat{(n+1)}\cdots}_{g}\ub{\cdots}_{g'} 1^+2^-) \\
(2^-\ub{\cdots\widehat{n}^-}_{g}\ub{\cdots}_{g'} | \ub{\widehat{(n+1)}\cdots}_{g}\ub{\cdots}_{g'} 1^+) \\
(1^+2^-\ub{\cdots\widehat{n}^-}_{g}\ub{\cdots}_{g'} | \ub{\widehat{(n+1)}\cdots}_{g})
\end{split}
\end{align}
where $g$ denote the gluons of the gauge group including $n^-$ and $(n+1)^+$ and $g'$ the gluons of the other.
Again these three types belong to different number sets.
\\ \indent
Now suppose each of the above types of configurations is nonvanishing.
Then if we want to make other configurations from each configuration preserving the number set, we should perform irreducible OPP or overall cyclic permutation including $1^+$ and $2^-$. 
In fact, with these permutations we can, from one configuration in a number set, generate all the other configurations which belong to the same number set.
The other permutations give the configurations which belong to other color bases.
We see that, from each type of configurations, it is impossible to generate different nonvanishing configurations in the same number set. 
This means that each of the above types is, if nonvanishing, the only configuration in each number set.
\\ \indent
Let us examine this argument in detail with the first configuration of (\ref{eq:case1}).
Suppose we want to change the configuration into a different one which belongs to the same number set.
Then if we move a particle from the upper amplitude to the lower amplitude, we should move another particle from the lower amplitude to the upper amplitude to preserve the number set. 
Now, let us try to move one of $g'$ in the upper amplitude to the lower amplitude. 
Then from the lower amplitude we should move one of $g$, one of $g'$, or $1^+$ to the upper amplitude.
In the case of one of $g$, it is impossible to move the particle since the order of $g$ particles should be preserved in the OPP.  
It is also impossible to move one of $g'$, because to move it to the upper amplitude it is permuted with the one moved from the upper amplitude and this is not permitted in OPP.
Finally, we cannot move $1^+$ to the upper amplitude, since the resulting lower amplitude consists of only gluons of different gauge groups without $q\bar{q}$ pair and vanishes.
In result, it is impossible to make a different nonvanishing configuration without changing the number set. 
Note that the reducible OPP generates the same configuration.
\\ \indent
By similar arguments, we confirm that the other types of configurations in (\ref{eq:case1}) and (\ref{eq:case2}) are the only nonvanishing configurations in each number set.
In result, for the case of product gauge group amplitudes, each number set has at most two configurations. (Each type has two configurations allowing two helicity compositions of the internal line, as in the case of single gauge group amplitudes.) This means that, as asserted, the number of configurations for a product gauge group amplitude is equal to or smaller than that of a single amplitude.
\\ \indent
If we consider the configurations that do not contribute because of the helicity compositions, the situation may get better. 
In product gauge group amplitudes, all types of vanishing configurations owing to the helicity compositions in single gauge group amplitudes also vanish and there is another type of vanishing configuration such as $(2^-\cdots|^{-+}|\cdots 1^+)$, which appeared in section \ref{sub:config}. 
This type of configuration does not contribute since internal particle $\widehat{P}$ is a fermion and should have opposite helicity to $2^-$($1^+$) in the upper(lower) amplitude.

\section{7-Point Next-to-MHV Amplitudes} \label{7p}

We present here just the results of 7-point primitive amplitudes, $A_{34}$ and $A_{3}$, without explicit configurations. All results coincide with the sums over the single gauge group amplitudes as expected from (\ref{eq:singlesum}).
The 7-point single gauge group amplitudes used in the comparison are given in appendix \ref{7ps}.

\subsection{$A_{34}$}
\begin{align} 
%\begin{split}
A_{34}^{--+++} 
&= -\frac{\la 2|3+4|1]^2}
                {[34][41]\la 25\ra\la 56\ra\la 67\ra\la 7|1+4|3]} \nn \\
&       \quad +\frac{\la 2|(5+6)(7+1)|4\ra^2}
                {t_{147}\la 71\ra\la 25\ra\la 56\ra\la 6|5+2|3]\la 7|1+4|3]} \nn \\
& \quad +\frac{\la 4|3+2|5]^2}
                {t_{235}[23]\la 67\ra\la 71\ra\la 6|5+2|3]}
%\end{split} 
\\ \nn \\
%\begin{split}
A_{34}^{-+-++} 
&= \frac{\la 31\ra\la 5|6+7|2]\la 5|(6+7)(1+4)|3\ra^2}
                {t_{567}t_{134}\la 56\ra\la 67\ra\la 34\ra\la 41\ra\la 7|6+5|2]\la 1|4+3|2]} \nn \\
&       \quad +\frac{\la 31\ra[26]\la 3|2+5|6]^2}
                {t_{256}[25][56]\la 71\ra\la 34\ra\la 41\ra\la 7|6+5|2]} \nn \\
& \quad -\frac{\la 15\ra[24]\la 5|2+3|4]^2}
                {t_{234}[23][34]\la 56\ra\la 67\ra\la 71\ra\la 1|4+3|2]}
%\end{split} 
\\ \nn \\
%\begin{split}
A_{34}^{-++-+} 
&= -\frac{\la 61\ra\la 31\ra(\la 31\ra\la 6|2+3|5]+\la 63\ra\la 41\ra[45])^3}
                {\la 67\ra\la 71\ra\la 34\ra\la 41\ra\la 1|4+3|2]\la 1|7+6|5]\la 3|(2+5)(6+7)|1\ra\la 6|(2+5)(3+4)|1\ra} \nn \\
&       \quad +\frac{\la 31\ra\la 23\ra^2[57]^4}
                {t_{567}[56][67]\la 34\ra\la 41\ra\la 2|5+6|7]\la 1|7+6|5]} \nn \\
&       \quad -\frac{\la 31\ra\la 26\ra^3\la 3|4+1|7]^2\la 6|2+5|7]}
                {t_{256}t_{134}\la 25\ra\la 56\ra\la 34\ra\la 41\ra\la 2|5+6|7]\la 6|(2+5)(3+4)|1\ra} \nn \\
&       \quad +\frac{\la 16\ra\la 23\ra^2\la 6|7+1|4]^2\la 3|2+5|4]}
                {t_{671}t_{235}\la 25\ra\la 67\ra\la 71\ra\la 5|2+3|4]\la 3|(2+5)(6+7)|1\ra} \nn \\
&       \quad -\frac{\la 16\ra[24]\la 6|2+3|4]^3}
                {t_{234}[23][34]\la 56\ra\la 67\ra\la 71\ra\la 5|2+3|4]\la 1|4+3|2]}
%\end{split} 
\\ \nn \\
%\begin{split}
A_{34}^{-+++-} 
&= \frac{\la 31\ra\la 23\ra^2[56]^3}
                {t_{567}[67]\la 34\ra\la 41\ra\la 1|7+6|5]\la 2|5+6|7]} \nn \\
&       \quad -\frac{\la 31\ra\la 3|(1+4)(5+6)|2\ra^3}
                {t_{134}\la 25\ra\la 56\ra\la 34\ra\la 41\ra\la 3|4+1|7]\la 2|5+6|7]\la 6|(2+5)(3+4)|1\ra} \nn \\
& \quad -\frac{\la 31\ra(\la 31\ra\la 7|2+3|5]+ \la 73\ra\la 41\ra[45])^3}
                {\la 67\ra\la 34\ra\la 41\ra\la 1|4+3|2]\la 1|7+6|5]\la 3|(2+5)(6+7)|1\ra\la 6|(2+5)(3+4)|1\ra} \nn \\
& \quad +\frac{\la 23\ra^2[14]^2\la 3|7+1|4]}
                {t_{471}\la 25\ra\la 56\ra[71]\la 6|7+1|4]\la 3|4+1|7]} \nn \\
& \quad -\frac{\la 23\ra^2\la 7|6+1|4]^3\la 3|2+5|4]}
                {t_{671}t_{235}\la 25\ra\la 67\ra\la 5|2+3|4]\la 6|7+1|4]\la 3|(2+5)(6+7)|1\ra} \nn \\
& \quad +\frac{[24]\la 7|2+3|4]^3}
                {t_{234}[23][34]\la 56\ra\la 67]\la 1|4+3|2]\la 5|2+3|4]}
%\end{split} 
\\ \nn \\
%\begin{split}
A_{34}^{+--++} 
&= -\frac{\la 24\ra^3\la 5|6+7|1]^2\la 5|(6+7)(2+3)|4\ra}
                {t_{234}t_{567}\la 23\ra\la 34\ra\la 56\ra\la 67\ra\la 2|3+4|1]\la 4|(2+3)(5+6)|7\ra} \nn \\
& \quad +\frac{\la 57\ra(\la 57\ra\la 4|5+2|3]+\la 45\ra\la 67\ra[63])^3}
                {\la 56\ra\la 67\ra\la 71\ra\la 7|6+5|2]\la 7|1+4|3]\la 7|(1+4)(2+3)|5]\ra\la 4|(2+3)(5+6)|7\ra} \nn \\
& \quad -\frac{\la 25\ra^2[13]^3\la 5|4+1|3]}
                {t_{134}[34][41]\la 56\ra\la 67\ra\la 7|1+4|3]\la 2|3+4|1]} \nn \\
& \quad +\frac{\la 25\ra^2\la 4|7+1|6]^2\la 5|2+3|6]}
                {t_{471}t_{235}\la 71\ra\la 23\ra\la 3|2+5|6]\la 7|(1+4)(2+3)|5\ra} \nn \\
& \quad +\frac{[26]\la 4|2+5|6]^3}
                {t_{256}[25][56]\la 34\ra\la 71\ra\la 3|2+5|6]\la 7|6+5|2]}
%\end{split} 
\\ \nn \\
%\begin{split}
A_{34}^{+-+-+} 
&= \frac{[17]^2\la 24\ra^3\la 26\ra^3(\la 26\ra\la 4|3+2|7]+\la 42\ra\la 56\ra[57])}
                {\la 23\ra\la 34\ra\la 25\ra\la 56\ra\la 2|5+6|7]\la 2|3+4|1]\la 2|(3+4)(7+1)|6\ra\la 2|(5+6)(7+1)|4\ra} \nn \\
& \quad -\frac{\la 26\ra^3\la 4|1+7|3]^3}
                {t_{147}t_{256}\la 71\ra\la 25\ra\la 56\ra\la 7|1+4|3]\la 2|(5+6)(7+1)|4\ra} \nn \\
& \quad -\frac{[13]^3\la 26\ra^3\la 6|1+4|3]}
                {t_{134}[34][41]\la 25\ra\la 56\ra\la 67\ra\la 7|4+1|3]\la 2|3+4|1]} \nn \\
& \quad +\frac{\la 24\ra^3[57]^4}{t_{567}\la 23\ra\la 34\ra[56][67]\la 1|7+6|5]\la 2|5+6|7]} \nn \\
& \quad +\frac{\la 16\ra\la 24\ra^3\la 6|7+1|5]^3}
                {t_{671}t_{234}\la 23\ra\la 34\ra\la 67\ra\la 71\ra\la 1|7+6|5]\la 2|(3+4)(7+1)|6\ra}
%\end{split} 
\\ \nn \\
%\begin{split}
A_{34}^{+-++-} 
&= -\frac{\la 24\ra^3\la 2|5+6|1]^3}
                {\la 23\ra\la 34\ra\la 25\ra\la 56\ra[71]\la 2|5+6|7]\la 2|3+4|1]\la 2|(3+4)(7+1)|6\ra} \nn \\
& \quad -\frac{\la 27\ra^3[13]^3}
                {t_{341}[34][41]\la 25\ra\la 56\ra\la 67\ra\la 2|3+4|1]} \nn \\
& \quad +\frac{\la 24\ra^3[56]^3}
                {t_{567}\la 23\ra\la 34\ra[67]\la 1|7+6|5]\la 2|5+6|7]} \nn \\
& \quad -\frac{\la 24\ra^3\la 7|1+6|5]^3}
                {t_{671}t_{234}\la 23\ra\la 34\ra\la 67\ra\la 1|7+6|5]\la 2|(3+4)(7+1)|6\ra}
%\end{split} 
\\ \nn \\
%\begin{split}
A_{34}^{++--+} 
&= \frac{\la 61\ra\la 6|(7+1)(3+4)|2\ra^2}
                {t_{671}\la 23\ra\la 34\ra\la 67\ra\la 71\ra\la 1|7+6|5]\la 4|3+2|5]} \nn \\
& \quad +\frac{\la 61\ra\la 6|5+2|3]^2}
                {t_{235}\la 41\ra\la 67\ra\la 71\ra[25]\la 4|3+2|5]} \nn \\
& \quad -\frac{\la 2|5+6|7]^2\la 1|5+6|7]}
                {t_{567}[56][67]\la 23\ra\la 34\ra\la 41\ra\la 1|7+6|5]}
%\end{split} 
\\ \nn \\
%\begin{split}
A_{34}^{++-+-} 
&= \frac{\la 57\ra^4\la 2|3+4|1]^2}
                {t_{567}\la 23\ra\la 34\ra\la 56\ra\la 67\ra\la 5|6+7|1]\la 4|(2+3)(5+6)|7\ra} \nn \\
& \quad -\frac{\la 57\ra^4\la 7|1+4|3]^2}
                {\la 41\ra\la 56\ra\la 67\ra\la 7|6+5|2]\la 7|(1+4)(2+3)|5\ra\la 4|(2+3)(5+6)|7\ra} \nn \\
& \quad -\frac{\la 25\ra^2[16]^3\la 5|7+1|6]}
                {t_{671}\la 23\ra\la 34\ra[67][71]\la 4|7+1|6]\la 5|6+7|1]} \nn \\
& \quad +\frac{\la 25\ra^2\la 7|4+1|6]^3\la 5|2+3|6]
                }{t_{471}t_{235}\la 23\ra\la 41\ra\la 3|2+5|6]\la 4|7+1|6]\la 7|(1+4)(2+3)|5\ra} \nn \\
& \quad +\frac{[26]\la 7|2+5|6]^3}
                {t_{256}[25][56]\la 34\ra\la 41\ra\la 3|2+5|6]\la 7|6+5|2]} 
%\end{split} 
\\ \nn \\
%\begin{split}
A_{34}^{+++--} 
&= -\frac{\la 26\ra^3\la 2|3+4|1]^2}
                {\la 23\ra\la 34\ra\la 25\ra\la 56\ra[71]\la 2|5+6|7]\la 2|(5+6)(1+7)|4\ra} \nn \\
& \quad -\frac{\la 26\ra^3\la 7|1+4|3]^2}
                {t_{256}t_{714}\la 41\ra\la 25\ra\la 56\ra\la 2|(5+6)(1+7)|4\ra} \nn \\
& \quad -\frac{\la 2|7+6|5]^3}
                {t_{567}\la 23\ra\la 34\ra\la 41\ra[56][67]\la 2|5+6|7]}
%\end{split} 
\end{align}

\subsection{$A_3$}

\begin{align} 
%\begin{split}
A_3^{--+++}
&= \frac{\la 4|3+2|1]^2\la 4|3+1|2]}{t_{123}[23][31]\la 45\ra\la 56\ra\la 67\ra\la 7|1+3|2]} \nn \\
& \quad +\frac{\la 4|5+6|2]\la 4|(5+6)(7+1)|3]^2}{t_{456}t_{713}\la 71\ra\la 45\ra\la 56\ra\la 6|5+4|2]\la 7|1+3|2]} \nn \\
& \quad +\frac{[25]\la 3|2+4|5]^2}{t_{245}[24][45]\la 67\ra\la 71\ra\la 6|5+4|2]}
%\end{split} 
\\ \nn \\
%\begin{split}
A_3^{-+-++}
&= \frac{\la 51\ra\la 5|2+3|4]^3}
                {t_{234}[23]\la 56\ra\la 67\ra\la 71\ra\la 5|2+4|3]\la 1|2+3|4]} \nn \\
& \quad +\frac{\la 23\ra^2\la 5|6+7|4]^4}
                {t_{567}t_{123}\la 56\ra\la 67\ra\la 76+54\ra\la 1|2+3|4]\la 5|(6+7)(1+3)|2\ra} \nn \\
& \quad +\frac{\la 25\ra^3\la 5|6+7|1]^2\la 5|(1+3)(6+7)|5\ra}
                {\la 24\ra\la 45\ra\la 56\ra\la 67\ra[31]\la 5|2+4|3]\la 7|(1+3)(2+4)|5\ra\la 2|(1+3)(6+7)|5\ra} \nn \\
& \quad +\frac{\la 25\ra^3\la 3|7+1|6]^2\la 5|2+4|6]}
                {t_{245}t_{713}\la 24\ra\la 45\ra\la 71\ra\la 2|4+5|6]\ra 7|(1+3)(2+4_|5\ra} \nn \\
& \quad +\frac{\la 23\ra^2[46]^4}
                {t_{456}\la 71\ra[45][56]\la 2|4+5|6]\la 7|6+5|4]} 
%\end{split} 
\\ \nn \\
%\begin{split}
A_3^{-++-+}
&= \frac{\la 23\ra^2[45]^3\la 61\ra^4}
                {\la 67\ra\la 71\ra\la 1|7+6|5]\la 1|2+3|4]\la 2|(4+5)(6+7)|1\ra\la 1|(2+3)(4+5)|6\ra} \nn \\
& \quad -\frac{\la 61\ra\la 6|(7+1)(4+5)|2\ra^3}
                {t_{671}\la 24\ra\la 45\ra\la 67\ra\la 71\ra\la 6|7+1|3]\la 5|2+4|3]\la 2|(4+5)(6+7)|1\ra} \nn \\
& \quad +\frac{\la 61\ra\la 6|2+3|4]^3}
                {t_{234}[23]\la 56\ra\la 67\ra\la 71\ra\la 5|2+4|3]\la 1|2+3|4]} \nn \\
& \quad +\frac{\la 26\ra^3[17]^2\la 6|1+3|7]}
                {t_{713}\la 24\ra\la 45\ra\la 56\ra[31]\la 2|1+3|7]\la 6|7+1|3]} \nn \\
& \quad +\frac{\la 23\ra^2\la 6|5+4|7]^4}
                {t_{123}t_{456}\la 45\ra\la 56\ra\la 4|5+6|7]\la 2|1+3|7]\la 1|(2+3)(4+5)|6\ra} \nn \\
& \quad +\frac{\la 23\ra^2[57]^4}
                {t_{567}[56][67]\la 24\ra\la 4|5+6|7]\la 1|7+6|5]}
%\end{split} 
\\ \nn \\
%\begin{split}
A_3^{-+++-}
&= \frac{\la 23\ra^2\la 2|4+5|6]^3}
                {[67]\la 24\ra\la 45\ra\la 2|3+1|7]\la 5|(6+7)(1+3)|2\ra\la 2|(4+5)(6+7)|1\ra} \nn \\
& \quad +\frac{\la 2|7+3|1]^3}
                {\la 24\ra\la 45\ra\la 56\ra[71][13]\la 6|7+1|3]\la 2|3+1|7]} \nn \\
& \quad +\frac{\la 2|(4+5)(6+1)|7\ra^3}
                {t_{671}\la 24\ra\la 45\ra\la 67\ra\la 5|4+2|3]\la 6|7+1|3]\la 2|(4+5)(6+7)|1\ra} \nn \\
& \quad +\frac{\la 23\ra^2\la 7|6+5|4]^3}
                {t_{567}t_{123}\la 56\ra\la 67\ra\la 1|2+3|4]\la 5|(6+7)(1+3)|2\ra} \nn \\
& \quad +\frac{\la 7|2+3|4]^3}
                {t_{234}[23]\la 56\ra\la 67\ra\la 5|4+2|3]\la 1|2+3|4]}
%\end{split} 
\\ \nn \\
%\begin{split}
A_3^{+--++}
&= \frac{\la 51\ra\la 5|2+4|3]^2}
                {t_{234}[24]\la 56\ra\la 67\ra\la 71\ra\la 1|2+3|4]} \nn \\
& \quad -\frac{\la 5|(6+7)(3+2)|1\ra\la 5|(6+7)(3+1)|2\ra^2}
                {t_{567}t_{123}\la 23\ra\la 3\ra\la 56\ra\la 67\ra\la 7|6+5|4]\la 1|2+3|4]} \nn \\
& \quad -\frac{\la 2|4+5|6]^2\la 1|4+5|6]}
                {t_{456}[45][56]\la 23\ra\la 31\ra\la 71\ra\la 7|6+5|4]} 
%\end{split} 
\\ \nn \\
%\begin{split}
A_3^{+-+-+}
&= \frac{\la 16\ra\la 46\ra^4\la 6|7+1|3]^2}
                {\la 45\ra\la 56\ra\la 67\ra\la 71\ra\la 6|5+4|2]\la 4|(3+2)(1+7)|6\ra\la 6|(5+4)(3+2)|1\ra} \nn \\
& \quad +\frac{\la 16\ra\la 24\ra^2\la 6|7+1|5]^3\la 4|3+2|5]}
                {t_{671}t_{234}\la 23\ra\la 67\ra\la 71\ra\la 1|7+6|5]\la 3|2+4|5]\la 4|(3+2)(1+7)|6\ra} \nn \\
& \quad -\frac{\la 16\ra[25]\la 6|2+4|5]^3}
                {t_{245}[24][45]\la 67\ra\la 71\ra\la 31\ra\la 3|2+4|5]\la 6|5+4|2]} \nn \\
& \quad -\frac{\la 46\ra^4\la 2|1+3|7]^2\la 1|2+3|7]}
                {t_{123}t_{456}\la 23\ra\la 31\ra\la 45\ra\la 56\ra\la 4|5+6|7]\la 1|(2+3)(4+5)|6\ra} \nn \\
& \quad -\frac{\la 14\ra\la 24\ra^2[57]^4}
                {t_{567}\la 23\ra\la 31\ra[56][67]\la 4|5+6|7]\la 1|7+6|5]}
%\end{split} 
\\ \nn \\
%\begin{split}
A_3^{+-++-}
&= -\frac{[13]^2\la 47\ra^4}
                {t_{123}\la 45\ra\la 56\ra\la 67\ra\la 7|1+3|2]\la 4|3+2|1]} \nn \\
& \quad -\frac{\la 24\ra^2\la 4|5+6|1]^3\la 4|(5+6)(7+1)|4\ra}
                {\la 23\ra\la 45\ra\la 56\ra[71]\la 4|5+6|7]\la 4|3+2|1]\la 4|(3+2)(1+7)|6\ra\la 4|(5+6)(7+1)|3\ra} \nn \\
& \quad -\frac{\la 4|5+6|2]\la 7|(1+3)(6+5)|4\ra^3}
                {t_{456}t_{713}\la 45\ra\la 56\ra\la 31\ra\la 6|5+4|2]\la 7|1+3|2]\la 4|(5+6)(7+1)|3\ra} \nn \\
& \quad -\frac{\la 24\ra^2\la 14\ra[56]^3}
                {t_{567}\la 23\ra\la 31\ra[67]\la 1|7+6|5]\la 4|5+6|7]} \nn \\
& \quad +\frac{\la 24\ra^2\la 7|6+1|5]^3\la 4|2+3|5]}
                {t_{671}t_{234}\la 23\ra\la 67\ra\la 3|2+4|5]\la 1|7+6|5]\la 6|(7+1)(2+3)|4\ra} \nn \\
& \quad +\frac{[25]\la 7|2+4|5]^3}
                {t_{245}[24][45]\la 31\ra\la 67\ra\la 32+45]\la 6|5+4|2]}
%\end{split} 
\\ \nn \\
%\begin{split}
A_3^{++--+}
&= \frac{\la 61\ra(\la 61\ra\la 2|5+6|4]+\la 26\ra\la 71\ra[74])^3}
                {\la 23\ra\la 31\ra\la 67\ra\la 71\ra[45]\la 1|7+6|5]\la 1|(2+3)(4+5)|6\ra\la 2|(4+5)(6+7)|1\ra} \nn \\
& \quad -\frac{\la 61\ra\la 25\ra^3\la 6|7+1|3]^2}
                {t_{671}t_{245}\la 24\ra\la 45\ra\la 67\ra\la 71\ra\la 2|(4+5)(6+7)|1\ra} \nn \\
& \quad -\frac{\la 56\ra^3\la 2|1+3|7]^2\la 1|2+3|7]}
                {t_{123}t_{456}\la 45\ra\la 23\ra\la 31\ra\la 4|5+6|7]\la 1|(2+3)(4+5)|6\ra} \nn \\
& \quad -\frac{\la 2|5+6|7]^3\la 1|5+6|7]}
                {t_{567}\la 23\ra\la 31\ra\la 24\ra[56][67]\la 4|5+6|7]\la 1|7+6|5]}
%\end{split} 
\\ \nn \\
%\begin{split}
A_3^{++-+-}
&= -\frac{\la 57\ra^4\la 2|1+3|4]^3}
                {t_{567}t_{123}\la 23\ra\la 31\ra\la 56\ra\la 67\ra\la 7|6+5|4]\la 5|(6+7)(1+3)|2\ra} \nn \\
& \quad -\frac{[13]^2\la 25\ra^3\la 57\ra^4}
                {\la 24\ra\la 45\ra\la 56\ra\la 67\ra\la 5|6+7|1]\la 7|(1+3)(2+4)|5\ra\la 5|(6+7)(1+3)|2\ra} \nn \\
& \quad -\frac{\la 25\ra^3[16]^3\la 5|7+1|6]}
                {t_{671}\la 23\ra\la 24\ra\la 45\ra[67][71]\la 3|7+1|6]\la 5|6+7|1]} \nn \\
& \quad +\frac{\la 25\ra^3\la 7|1+3|6]^3\la 5|2+4|6]}
                {t_{245}t_{713}\la 24\ra\la 45\ra\la 31\ra\la 2|4+5|6]\la 3|7+1|6]\la 7|(1+3)(2+4)|5\ra} \nn \\
& \quad +\frac{\la 27\ra^3[46]^4}
                {t_{456}\la 23\ra\la 31\ra[45][56]\la 2|4+5|6]\la 7|6+5|4]}
%\end{split} 
\\ \nn \\
%\begin{split}
A_3^{+++--}
&= -\frac{\la 2|(4+5)(6+7)|2\ra^3}
                {\la 23\ra\la 31\ra\la 24\ra\la 45\ra[67]\la 2|4+5|6]\la 2|1+3|7]\la 5|(6+7)(1+3)|2\ra} \nn \\
& \quad +\frac{[13]^2\la 26\ra^3}
                {t_{137}[71]\la 24\ra\la 45\ra\la 56\ra\la 2|1+3|7]} \nn \\
& \quad +\frac{\la 27\ra^3[45]^3}
                {t_{456}\la 23\ra\la 31\ra[56]\la 7|6+5|4]\la 2|4+5|6]} \nn \\
& \quad -\frac{\la 67\ra^3\la 2|1+3|4]^3}
                {t_{567}t_{123}\la 23\ra\la 31\ra\la 56\ra\la 7|6+5|4]\la 5|(6+7)(1+3)|2\ra}
%\end{split} 
\end{align}

\section{7-Point $A_S$} \label{7ps}
We provide all 7-point next-to-MHV single gauge group amplitudes, which are used in the comparison with the results of the 7-point product gauge group amplitudes.
\begin{align} 
%\begin{split}
A_S^{--+++}
&= -\frac{\la 4|3+2|1]^2\la 4|3+1|2]}
                {t_{123}[12][23]\la 45\ra\la 56\ra\la 67\ra\la 7|1+2|3]} \nn \\
& \quad +\frac{\la 4|(5+6)(7+1)|2\ra^2\la 4|(5+6)(7+2)|1\ra}
    {t_{456}t_{712}\la 71\ra\la 12\ra\la 45\ra\la 56\ra\la 6|5+4|3]\la 7|1+2|3]} \nn \\
& \quad +\frac{\la 2|3+4|5]^2\la 1|3+4|5]}
                {t_{345}[34][45]\la 67\ra\la 71\ra\la 12\ra\la 6|5+4|3]}
%\end{split} 
\\ \nn \\
%\begin{split}
A_S^{-+-++}
&= -\frac{\la 35\ra^4\la 5|6+7|1]^2\la 5|6+7|2]}
                {[12]\la 34\ra\la 45\ra\la 56\ra\la 67\ra\la 5|4+3|2]\la 5|(6+7)(1+2)|3\ra\la 7|(1+2)(3+4)|5\ra} \nn \\
& \quad +\frac{\la 35\ra^4\la 2|1+7|6]^2\la 1|2+7|6]}
    {t_{712}t_{345}\la 71\ra\la 12\ra\la 34\ra\la 45\ra\la 3|4+5|6]\la 7|(1+2)(3+4)|5\ra} \nn \\
& \quad +\frac{\la 13\ra\la 23\ra^2[46]^4}
                {t_{456}\la 71\ra\la 12\ra[45][56]\la 7|6+5|4]\la 3|4+5|6]} \nn \\
& \quad -\frac{\la 13\ra\la 23\ra^2\la 5|6+7|4]^4}
    {t_{567}t_{123}\la 12\ra\la 56\ra\la 67\ra\la 1|2+3|4]\la 7|6+5|4]\la 3|(2+1)(7+6)|5\ra} \nn \\
& \quad -\frac{\la 15\ra[24]\la 5|2+3|4]^3}
                {t_{234}[23][34]\la 56\ra\la 67\ra\la 71\ra\la 5|4+3|2]\la 1|2+3|4]}
%\end{split} 
\\ \nn \\
%\begin{split}
A_S^{-++-+}
&= -\frac{\la 31\ra\la 32\ra^2\la 3|4+5|7]^4}
                {\la 12\ra\la 34\ra\la 45\ra[67]\la 3|4+5|6]\la 3|2+1|7]\la 5|(6+7)(1+2)|3\ra\la 3|(4+5)(6+7)|1\ra} \nn \\
& \quad -\frac{\la 16\ra\la 34+52]\la 3|(4+5)(1+7)|6\ra^3}
    {t_{345}t_{671}\la 17\ra\la 76\ra\la 34\ra\la 45\ra\la 5|4+3|2]\la 6|7+1|2]\la 3|(4+5)(6+7)|1\ra} \nn \\
& \quad +\frac{\la 36\ra^4[17]^2[27]}
                {t_{712}\la 34\ra\la 45\ra\la 56\ra[12]\la 3|2+1|7]\la 6|7+1|2]} \nn \\
& \quad +\frac{\la 13\ra\la 23\ra^2[45]^3}
                {t_{456}\la 71\ra\la 12\ra[56]\la 7|6+5|4]\la 3|4+5|6]} \nn \\
& \quad -\frac{\la 13\ra\la 23\ra^2\la 6|5+7|4]^4}
    {t_{567}t_{123}\la 12\ra\la 56\ra\la 67\ra\la 1|2+3|4]\la 7|6+5|4]\la 3|(2+1)(7+6)|5\ra} \nn \\
& \quad -\frac{\la 16\ra[24]\la 6|2+|34]^3}
                {t_{234}[23][34]\la 56\ra\la 67\ra\la 71\ra\la 5|4+3|2]\la 1|2+3|4]}
%\end{split} 
\\ \nn \\
%\begin{split}
A_S^{-+++-}
&= -\frac{\la 3|7+2|1]^3\la 3|7+1|2]}
                {t_{712}[71][12]\la 34\ra\la 45\ra\la 56\ra\la 3|2+1|7]\la 6|7+1|2]} \nn \\
& \quad -\frac{\la 13\ra\la 23\ra^2\la 3|4+5|6]^3}
                {\la 12\ra\la 34\ra\la 45\ra[67]\la 3|2+1|7]\la 5|(6+7)(1+2)|3\ra\la 1|(7+6)(5+4)|3\ra} \nn \\
& \quad +\frac{\la 3|4+5|2]\la 7|(6+1)(5+4)|3\ra^3}
    {t_{345}t_{671}\la 34\ra\la 45\ra\la 67\ra\la 5|4+3|2]\la 6|7+1|2]\la 1|(7+6)(5+4)|3\ra} \nn \\
& \quad +\frac{\la 13\ra\la 23\ra^2\la 7|6+|54]^3}
                {t_{567}t_{123}\la 12\ra\la 56\ra\la 67\ra\la 1|2+3|4]\la 5|(6+7)(1+2)|3\ra} \nn \\
& \quad +\frac{[24]\la 7|2+3|4]^3}
                {t_{234}[23][34]\la 56\ra\la 67\ra\la 5|4+3|2]\la 1|2+3|4]}
%\end{split} 
\\ \nn \\
%\begin{split}
A_S^{+--++}
&= -\frac{\la 15\ra\la 5|4+2|3]^3}
                {t_{234}[34]\la 56\ra\la 67\ra\la 71\ra\la 1|2+3|4]\la 5|4+3|2]} \nn \\
& \quad -\frac{\la 45\ra^3\la 5|6+7|1]^2\la 5|6+7|2]}
                {[12]\la 34\ra\la 56\ra\la 67\ra\la 5|4+3|2]\la 7|(1+2)(3+4)|5\ra\la 5|(6+7)(1+2)|3\ra} \nn \\
& \quad +\frac{\la 5|(6+7)(3+2)|1\ra\la 5|(6+7)(3+1)|2\ra^3}
    {t_{567}t_{123}\la 12\ra\la 23\ra\la 56\ra\la 67\ra\la 7|6+5|4]\la 1|2+3|4]\la 5|(6+7)(1+2)|3\ra} \nn \\
& \quad +\frac{\la 45\ra^3\la 2|1+7|6]^2\la 1|2+7|6]}
                {t_{712}t_{345}\la 71\ra\la 12\ra\la 34\ra\la 3|4+5|6]\la 7|(1+2)(3+4)|5\ra} \nn \\
& \quad +\frac{\la 2|4+5|6]^3\la 1|4+5|6]}
                {t_{456}[45][56]\la 71\ra\la 12\ra\la 23\ra\la 7|6+5|4]\la 3|4+5|6]}
%\end{split} 
\\ \nn \\
%\begin{split}
A_S^{+-+-+}
&= \frac{\la 14\ra\la 24\ra^3[57]^4}
                {t_{567}\la 12\ra\la 23\ra\la 34\ra[56][67]\la 4|5+6|7]\la 1|7+6|5]} \nn \\
& \quad -\frac{\la 16\ra\la 24\ra^3\la 4|3+2|5]\la 6|7+1|5]^3}
    {t_{671}t_{234}\la 23\ra\la 34\ra\la 67\ra\la 71\ra\la 2|3+4|5]\la 1|7+6|5]\la 6|(7+1)(2+3)|4\ra} \nn \\
& \quad -\frac{\la 24\ra^3[71]^2\la 46\ra^4\la 4|3+2|7]}
    {\la 23\ra\la 34\ra\la 45\ra\la 56\ra\la 4|3+2|1]\la 4|5+6|7]\la 4|(5+6)(7+1)|2\ra\la 6|(7+1)(2+3)|4\ra} \nn \\
& \quad +\frac{\la 16\ra\la 26\ra^3[35]^4}
                {t_{345}\la 67\ra\la 71\ra\la 12\ra[34][45]\la 6|5+4|3]\la 2|3+4|5]} \nn \\
& \quad +\frac{\la 46\ra^4\la 1|2+7|3]\la 2|1+7|3]^3}
    {t_{456}t_{712}\la 71\ra\la 12\ra\la 45\ra\la 56\ra\la 7|1+2|3]\la 6|5+4|3]\la 4|(5+6)(7+1)|2\ra} \nn \\
& \quad -\frac{[13]^3\la 46\ra^4}
                {t_{123}[12]\la 45\ra\la 56\ra\la 67\ra\la 4|3+2|1]\la 7|1+2|3]}
%\end{split} 
\\ \nn \\
%\begin{split}
A_S^{+-++-}
&= -\frac{[13]^3\la 47\ra^4}
                {t_{123}[12]\la 45\ra\la 56\ra\la 67\ra\la 7|1+2|3]\la 4|3+2|1]} \nn \\
& \quad -\frac{\la 72\ra^3\la 4|5+6|3]^4}
                {t_{456}t_{712}\la 12\ra\la 45\ra\la 56\ra\la 6|5+4|3]\la 7|1+2|3]\la 2|(1+7)(6+5)|4\ra} \nn \\
& \quad +\frac{\la 42\ra^3\la 4|5+6|1]^3\la 4|(5+6)(7+1)|4\ra}
    {\la 23\ra\la 34\ra\la 45\ra\la 56\ra[71]\la 4|5+6|7]\la 4|3+2|1]\la 6|(7+1)(2+3)|4\ra\la 2|(1+7)(6+5)|4\ra} \nn \\
& \quad +\frac{\la 24\ra^3\la 14\ra[56]^3}
                {t_{567}\la 12\ra\la 23\ra\la 34\ra[67]\la 1|7+6|5]\la 4|5+6|7]} \nn \\
& \quad +\frac{\la 24\ra^3\la 4|2+3|5]\la 7|1+6|5]^3}
    {t_{671}t_{234}\la 23\ra\la 34\ra\la 67\ra\la 23+4|5]\la 1|7+6|5]\la 6|(7+1)(2+3)|4\ra} \nn \\
& \quad -\frac{[35]^4\la 27\ra^3}
                {t_{345}\la 67\ra\la 12\ra[34][45]\la 6|5+4|3]\la 2|3+4|5]}
%\end{split} 
\\ \nn \\
%\begin{split}
A_S^{++--+}
&= \frac{\la 16\ra\la 2|(3+4)(7+1)|6\ra^3}
                {t_{671}\la 67\ra\la 71\ra\la 23\ra\la 34\ra\la 2|3+4|5]\la 1|7+6|5]\la 6|(7+1)(2+3)|4\ra} \nn \\
& \quad -\frac{\la 56\ra^3\la 24\ra^3[71]^2\la 4|3+2|7]}
    {\la 23\ra\la 34\ra\la 45\ra\la 43+21]\la 4|5+6|7]\la 6|(7+1)(2+3)|4]\la 4|(5+6)(7+1)|2\ra} \nn \\
& \quad +\frac{\la 1|5+6|7]\la 2|5+6|7]^3}
                {t_{567}\la 12\ra\la 23\ra\la 34\ra[56][67]\la 4|5+6|7]\la 1|7+6|5]} \nn \\
& \quad +\frac{\la 16\ra\la 26\ra^3[34]^3}
                {t_{345}\la 67\ra\la 71\ra\la 12\ra[45]\la 6|5+4|3]\la 2|3+4|5]} \nn \\
& \quad +\frac{\la 56\ra^3\la 1|7+2|3]\la 2|1+7|3]^3}
    {t_{456}t_{712}\la 71\ra\la 12\ra\la 45\ra\la 7|1+2|3]\la 6|5+4|3]\la 4|(5+6)(7+1)|2\ra} \nn \\
& \quad -\frac{[13]^3\la 56\ra^3}
                {t_{123}[12]\la 45\ra\la 67\ra\la 4|3+2|1]\la 7|1+2|3]}
%\end{split} 
\\ \nn \\
%\begin{split}
A_S^{++-+-}
&= \frac{\la 57\ra^4\la 2|3+4|1]^3}
    {t_{567}\la 56\ra\la 67\ra\la 23\ra\la 34\ra\la 4|3+2|1]\la 5|6+7|1]\la 2|(3+4)(5+6)|7\ra} \nn \\
& \quad +\frac{\la 72\ra^3\la 2|3+4|6]^4}
    {\la 12\ra\la 23\ra\la 34\ra[56]\la 2|3+4|5]\la 2|1+7|6]\la 45+6)(7+1)|2\ra\la 2|(3+4)(5+6)|7\ra} \nn \\
& \quad +\frac{\la 52\ra^3[16]^3\la 5|1+7|6]}
    {t_{671}\la 23\ra\la 34\ra\la 45\ra[67][71]\la 2|1+7|6]\la 5|6+7|1]} \nn \\
& \quad -\frac{\la 27\ra^3[34]^3}
    {t_{345}\la 67\ra\la 12\ra[45]\la 6|5+4|3]\la 2|3+4|5]} \nn \\
& \quad -\frac{\la 27\ra^3\la 5|6+4|3]^4}
    {t_{456}t_{712}\la 12\ra\la 45\ra\la 56\ra\la 7|1+2|3]\la 6|5+4|3]\la 4|(5+6)(7+1)|2\ra} \nn \\
& \quad -\frac{[13]^3\la 57\ra^4}
    {t_{123}[12]\la 45\ra\la 56\ra\la 67\ra\la 4|3+2|1]\la 7|1+2|3]}
%\end{split} 
\\ \nn \\
%\begin{split}
A_S^{+++--}
&= -\frac{\la 2|6+7|1]^3}
    {\la 23\ra\la 34\ra\la 45\ra[67][71]\la 2|1+7|6]\la 5|6+7|1]} \nn \\
& \quad +\frac{\la 72\ra^3\la 2|3+4|5]^3}
    {\la 12\ra\la 23\ra\la 34\ra[56]\la 2|1+7|6]\la 4|(5+6)(7+1)|2\ra\la 2|(3+4)(5+6)|7\ra} \nn \\
& \quad +\frac{\la 67\ra^3\la 2|3+4|1]^3}
    {t_{567}\la 23\ra\la 34\ra\la 56\ra\la 4|3+2|1]\la 5|6+7|1]\la 2|(3+4)(5+6)|7\ra} \nn \\
& \quad -\frac{\la 27\ra^3\la 6|5+4|3]^3}
    {t_{456}t_{712}\la 12\ra\la 45\ra\la 56\ra\la 7|1+2|3]\la 4|(5+6)(7+1)|2\ra} \nn \\
& \quad -\frac{[13]^3\la 67\ra^3}
    {t_{123}[12]\la 45\ra\la 56\ra\la 4|3+2|1]\la 7|1+2|3]}
%\end{split} 
\end{align}

\end{document}